\documentclass[final,3p,times]{elsarticle}

\usepackage{lineno,hyperref}
\usepackage{lineno,hyperref}
\usepackage{graphicx}
\usepackage{dcolumn}
\usepackage{bm}
\usepackage{float}
\usepackage{epsfig}
\usepackage{booktabs}
\usepackage{subfigure}
\usepackage{graphics}
\usepackage{amssymb}
\usepackage{amsmath}
\usepackage{array}
\usepackage{color}
\usepackage{booktabs}
\usepackage{multirow}
\usepackage{setspace}
\usepackage{nomencl}

\makenomenclature

\usepackage{graphicx}

\modulolinenumbers[5]

\journal{Computers \& Mathematics with Applications}

\begin{document}

\begin{frontmatter}

\title{Discrete effects on some boundary schemes of multiple-relaxation-time lattice
Boltzmann model for convection-diffusion equations}

\author[myfirstaddress,mymainaddress]{Yao Wu}
\author[myfirstaddress,mymainaddress]{Yong Zhao}
\author[myfirstaddress,mymainaddress]{Zhenhua Chai}

\author[myfirstaddress,mymainaddress]{Baochang Shi\corref{mycorrespondingauthor}}
\cortext[mycorrespondingauthor]{Corresponding author}
\ead{shibc@hust.edu.cn}

\address[myfirstaddress]{School of Mathematics and Statistics, Huazhong University of Science and Technology, Wuhan 430074, China}
\address[mymainaddress]{Hubei Key Laboratory of Engineering Modeling and Scientific Computing, Huazhong University of Science and Technology, Wuhan 430074, China}

\begin{abstract}
In this paper, we perform a more general analysis on the discrete effects of some boundary schemes of the popular one- to three-dimensional D$n$Q$q$ multiple-relaxation-time lattice Boltzmann model for convection-diffusion equation (CDE). Investigated boundary schemes include anti-bounce-back(ABB) boundary scheme, bounce-back(BB) boundary scheme and non-equilibrium extrapolation(NEE) boundary scheme.
In the analysis, we adopt a transform matrix $\textbf{M}$ constructed by natural moments in the evolution equation, and the result of ABB boundary scheme is consistent with the existing work of orthogonal matrix $\textbf{M}$.
We also find that the discrete effect does not rely on the choice of transform matrix, and obtain a relation to determine some of the relaxation-time parameters which can be used to eliminate the numerical slip completely under some assumptions.
In this relation, the weight coefficient is considered as an adjustable parameter which makes the parameter adjustment more flexible. The relaxation factors associated with second moments can be used to eliminate the numerical slip of ABB boundary scheme and BB boundary scheme while the numerical slip can not be eliminated of NEE boundary scheme.
Furthermore, we extend the relations to complex-valued CDE, several numerical examples are used to test the relations.
\end{abstract}


\end{frontmatter}

\linenumbers

\section{introduction}
In recent years, the lattice Boltzmann method (LBM)  has gained much attention, and has been wildly used in many fields \cite{Qian1992, Aidun2010, Guo2013, Zhang2011}.
The LBM has some distinct advantages over traditional methods in dealing with Navier-Stokes equations \cite{Qian1993, Luo2003, Zheng2006, Guo2007, wang2019lattice} and convection-diffusion equations (CDEs) \cite{Chopard2009, Perko2014, Servan2008, Huang2014, Ginzburg2012, Chai2018}.
One of the advantages of LBM is dealing with the complex boundary conditions in porous media \cite{Pan2006, Jeong2008, wang2019effects, Xuan2010, Hussain2015, chai2019lattice}. When we solve the macroscopic partial differential equation, there are always discrete errors in the numerical scheme, the boundary discrete effect exists between the real boundary condition and the numerical solution in the boundary point.

To our knowledge, the discrete effect of the bounce-back(BB) scheme was first discussed for the Poiseuille flow.
Ginzburg and Adler \cite{Ginzburg1994} first performed a boundary condition analysis for the face-centered-hypercubic lattice Boltzmann (LB) model applied to the Poiseuille flow and a plane stagnation flow.
After that, He $et$ $al.$ \cite{HeXY1997} analyzed the discrete effect of BB boundary scheme in the Bhatnagar-Gross-Krook (BGK) model,
and found that the relaxation time $\tau$ has a significant influence on the BB scheme for the no-slip boundary condition.
In a similar way, Guo $et$ $al.$ \cite{Guo2007pre} studied the existing discrete effect of the discrete Maxwell's diffuse-reflection (DMDR) scheme and the combined bounce-back/specular-reflection (CBBSR) scheme. Then, they simulated the Poiseuille flow in the slip flow regime with the multiple-relaxation-time (MRT) LB model, and found that the BGK model cannot yield correct results in this regime owing to the discrete effect \cite{Guo2008}.
Due to find that the boundary schemes considered in Refs. \cite{Guo2007pre,Guo2008} are nonlocal, they are not suitable for fluid flows in complex geometries, Chai $et$ $al.$ \cite{Chai2010} developed a local scheme combined halfway bounce-back boundary condition and full diffusive boundary condition for microscale gas flows in complex geometries, and illustrated that to realize the exact slip boundary condition, the discrete effect must be included and corrected.
Lu $et$ $al.$ \cite{LuJH2012} proposed an immerse boundary MRT LB model, and presented a special relaxation between two relaxation time parameters in which can reduce the numerical boundary slip effectively. Recently, Ren $et$ $al.$ \cite{Ren2016} analyzed the discrete effects in the DMDR and CBBSR schemes for the rectangular LBE, and presented a reasonable approach to overcome these discrete effects in these two schemes.

We noted that all of above works focus on the discrete effect of BB boundary scheme for fluid flows. Subsequently, there are also some works on the discrete effect of anti-bounce-back (ABB) boundary schemes for CDEs.
Zhang $et$ $al.$ \cite{ZhangT} presented a general ABB boundary scheme of the BGK model for CDEs.
They performed an analysis on the discrete effect of the ABB boundary condition, and suggested that there is a numerical slip related to the lattice size in the diffusion of Couette flow between solid walls, which cannot be eliminated in the BGK model.
Then, Cui $et$ $al.$ \cite{Cui} analyzed the ABB boundary condition of the MRT model for CDEs.
They presented a theoretical analysis on the discrete effect of the ABB boundary scheme
for the simple problems with a parabolic distribution in one direction, and observed that the numerical slip can be eliminated in the MRT LB model by choosing the free relaxation parameters properly.
However, the analysis is limited to some special MRT LB models, e.g., D2Q4, D2Q5, and D2Q9 model.
Recently, based on the two-relaxation-times(TRT) model, Ginzburg $et$ $al.$ \cite{Ginzburg2017} presented a more general relation between the two relaxation factors through equating the set of closure relations of the given boundary scheme to the Taylor expansion.
In this work, based on the existing works \cite{Cui}, we firstly conduct the discrete effect on the ABB boundary scheme
of the more general MRT model composed of the natural moments for CDEs, and then derived a relation with four parameters : the weight coefficient, the relaxation factors $s_1$ and $s_2$ associated with first and second moments and a model parameter $\theta$ for adjustment to elimate the numerical slip.
After that, we conduct the discrete effect on BB boundary condition and non-equilibrium extrapolation(NEE) boundary condition, observed that the discrete effect can be elimated when $s_1+s_2=2$ on BB boundary condition and can not be elimated on NEE boundary condition. Furthermore, we observed that the relations is applicable to both real- and complex-valued problems, and has a general expression from one to three dimensions.

The paper is organized as follows. In Sec. \uppercase\expandafter{\romannumeral2}, we introduce the MRT model composed of natural moments.
Then we derive the equivalent finite-difference scheme of the MRT model for CDEs, and discuss the discrete effects on the ABB, BB, NEE boundary conditions in Sec. \uppercase\expandafter{\romannumeral3}.
Numerical tests are performed in Sec. \uppercase\expandafter{\romannumeral4}.
Finally, we give a brief summary in Sec. \uppercase\expandafter{\romannumeral5}.
\section{MRT LB model for convection-diffusion equation}
Firstly, we introduce the MRT model composed of the natural moments for CDEs.
The \textit{n}-dimensional (\textit{n}D) CDEs can be written as
\begin{equation}\label{eq1}
{\partial _t}\phi  + \mathbf{\nabla}  \cdot(\phi\textbf{u}) = \mathbf{\nabla} \cdot ({D\mathbf{\nabla}\phi}) + R,
\end{equation}
where $\phi$ is a scalar function of position $\textbf{x}$ and time \emph{t}, $\mathbf{\nabla}$ is the gradient operator
with respect to the position $\textbf{x}$ in \emph{n} dimensions.
$D$ is the diffusion coefficient, $\textbf{u}$ is the convection velocity and $R$ is the source term.

The evolution equation of the MRT model with D$n$Q$q$ lattice for the CDE can be written as
\begin{equation}\label{eq2}
\begin{aligned}
Collision:&f_i(\textbf{x},t)^{+}=f_i(\textbf{x},t)-(\textbf{M}^{-1}\textbf{S}\textbf{M})_{ik}(f_k(\textbf{x},t)-f_k^{eq}(\textbf{x},t))+\delta_t[\textbf{M}^{-1}(\textbf{I}-\frac{\theta \textbf{S}}{2})\textbf{M}]_{ik}R_k,\\
Streaming:&f_i(\textbf{x}+\textbf{c}_i\delta_t,t+\delta_t)=f_i(\textbf{x},t)^{+}
\end{aligned}
\end{equation}
where $\delta_t$ is time step, $\textbf{I}$ is the identity matrix, and $\textbf{S}$ is a diagonal relaxation matrix with non-negative elements. The transformation matrix $\textbf{M}$ is composed of natural moments \cite{Karlin}.
$\theta$ is a real parameter, corresponding to the MRT model \cite{Cui} for $\theta=1$ and a scheme in Ref. \cite{Guo2008pre} for $\theta=0$, respectively.
$f_i(\textbf{x}, t)$ and $f_i^{eq}(\textbf{x}, t)$ are the distribution function and equilibrium distribution function (EDF) associated with the discrete velocity $\textbf{c}_i$ at position \textbf{x} and time $t$ respectively, and $f_i(\textbf{x},t)^{+}$ is the distribution function after collision. And to simplify the derivation, only the following linear EDF is considered here,
\begin{equation}\label{eq3}
f_i^{eq}(\textbf{x},t)=w_i\phi(1+\frac{\textbf{c}_i\cdot{\textbf{u}}}{c_s^2}),
\end{equation}
where $\omega_i$ is the weight coefficient, $c_s$ is the so-called lattice sound speed. $R_i$ is the discrete source term, and can be defined as
\begin{equation}\label{eq4}
R_i=\omega_iR.
\end{equation}

Firstly, for the D$1$Q$3$ model, the set of discrete velocities are $\textbf{c}=\{-1,0,1\}c$, where $c=\delta_x/\delta_t$ with $\delta_x$ being the lattice spacing.
The transformation matrix $\textbf{M}=(\textbf{c}_{ix}^m) (m = 0, 1, 2)$, which can be expressed as $\textbf{M}=\textbf{C}_{d}\textbf{M}_{0}$ \cite{Chai2016},
\begin{equation}\label{eq5}
  \textbf{M}_0=\left(
  \begin{array}{ccc}
    1 & 1 & 1 \\
    -1 & 0 & 1 \\
    1 & 0 & 1 \\
  \end{array}
\right).
\end{equation}
\begin{equation}\label{eq6}
\textbf{C}_d=diag(1,c,c^2),
\end{equation}
\begin{equation}\label{eq7}
\textbf{S}=diag(s_0,s_1,s_2).
\end{equation}
As for the D2Q9 model, the discrete velocities can be given by
\begin{equation}\label{eq8}
  \textbf{c}=\left(
  \begin{array}{ccccccccc}
    0 & 1 & 0 & -1& 0 & 1 & -1& -1& 1 \\
    0 & 0 & 1 & 0 & -1& 1 & 1 & -1& -1 \\
  \end{array}
\right)c,
\end{equation}
and the transformation matrix as $\textbf{M}=(\textbf{c}^m_{ix}\textbf{c}^n_{iy})=\textbf{C}_{d}\textbf{M}_{0}$, $(m, n = 0, 1, 2, m+n\leq2)$,
\begin{equation}\label{eq9}
  \textbf{M}_0=\left(
  \begin{array}{ccccccccc}
    1 & 1 & 1 & 1 & 1 & 1 & 1 & 1 & 1 \\
    0 & 1 & 0 & -1& 0 & 1 & -1& -1& 1 \\
    0 & 0 & 1 & 0 & -1& 1 & 1 & -1& -1 \\
    0 & 1 & 0 & 1 & 0 & 1 & 1 & 1 & 1 \\
    0 & 0 & 1 & 0 & 1 & 1 & 1 & 1 & 1 \\
    0 & 0 & 0 & 0 & 0 & 1 & -1 & 1 & -1 \\
    0 & 0 & 0 & 0 & 0 & 1 & 1& -1 & -1 \\
    0 & 0 & 0 & 0 & 0 & 1 & -1& -1& 1\\
    0 & 0 & 0 & 0 & 0 & 1 & 1 & 1 & 1 \\
  \end{array}
\right).
\end{equation}
\begin{equation}\label{eq10}
\textbf{C}_d=diag(1,c,c,c^2,c^2,c^2,c^3,c^3,c^4),
\end{equation}
\begin{equation}\label{eq11}
\textbf{S}=diag(s_0,s_1,s_1,s_2,s_2,s_2,s_3,s_3,s_4).
\end{equation}
In the present MRT model, the macroscopic variable $\phi$ should be computed by
\begin{equation}\label{eq12}
\phi=\sum_if_i+\frac{\theta R}{2}\delta t.
\end{equation}

\section{Discrete effects of some boundary schemes}
We now analyze the discrete effects of these boundary scheme in the framework of the MRT model for CDE.
For simplicity, we conducted an analysis of Dirichlet boundary conditions for the simple steady problems with a parabolic distribution in one direction.
\subsection{Equivalent difference equation of the MRT model}
Firstly, we consider the D$1$Q$3$ MRT model for one-dimensional steady problems with const $R$, and set the distribution function as $f_i^j=f_i(x_j)$, with $x_j$ being a discrete grid point.
To make the derivation easier to understand, we rewrite Eq. (\ref{eq2}) as
\begin{equation}\label{eq13}
f_i^j=\left\{
\begin{aligned}
{f}_i^{j,+} & ,\quad & i=0 \\
{f}_i^{j-1,+} &,\quad   & i=1 \\
{f}_i^{j+1,+} &,\quad   & i=-1
\end{aligned}
\right.
\end{equation}
where
\begin{equation}\label{eq14}
{f}_i^{j,+}=f_i(x_j,t)-(\textbf{M}^{-1}\textbf{S}\textbf{M})_{ik}(f_k({x}_j,t)-f_k^{eq}({x}_j,t))+\delta_t[\textbf{M}^{-1}(\textbf{I}-\frac{\theta \textbf{S}}{2})\textbf{M}]_{ik}R_k,i=0,1,-1.
\end{equation}
\begin{figure}[ht]
 \centering
\includegraphics[scale=0.7]{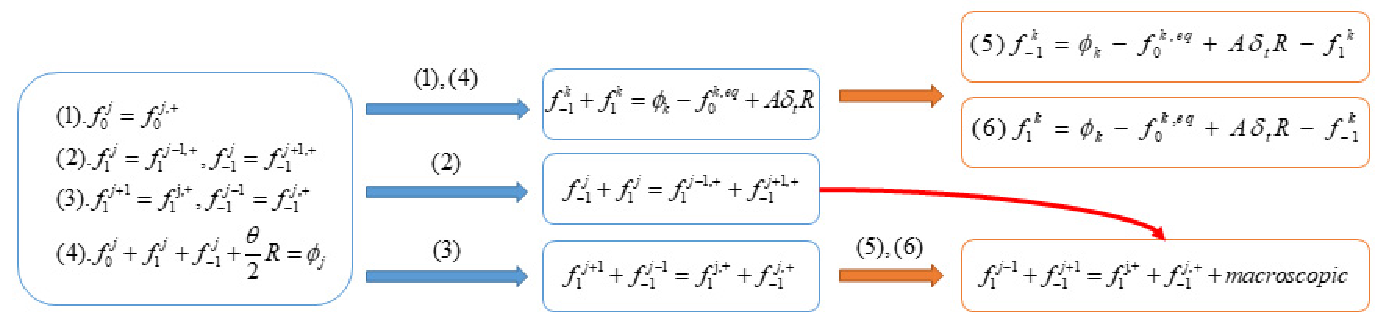}
\caption{The operation process to get the equivalent finite-difference scheme.} \label{Fig1}
\end{figure}
After taking some manipulations of the evolution equation, as shown in Fig. \ref{Fig1}(see Appendix A for details), we can obtain the following equivalent difference equation of the MRT model,
\begin{equation}\label{eq15}
\frac{\phi_{k+1}u_{k+1}-\phi_{k-1}u_{k-1}}{2\delta x}=D\frac{\phi_{k+1}-2\phi_k+\phi_{k-1}}{\delta x^2}+R,
\end{equation}
where $D=({1}/{s_1}-{1}/{2})c^2_s\delta t$, $c^2_s=2\omega_1c^2$.
Then we consider the D$2$Q$9$ MRT model for x-direction steady problems with constant $R$, and set the distribution function as $f_i^j=f_i(x_k,y_j)$, with $y_j$ being a discrete grid point,and $i$ being the direction of distribution function.
Eq. (\ref{eq2}) can be rewritten as
\begin{equation}\label{eq16}
f_i^j=\left\{
\begin{aligned}
{f}_i^{j,+} & ,\quad & i=0, 1, 3 \\
{f}_i^{j-1,+} &,\quad   & i=2, 5, 6 \\
{f}_i^{j+1,+} &,\quad   & i=4, 7, 8
\end{aligned}
\right.
\end{equation}
where ${f}_i^{j,+}=f_i(x_k,y_j,t)^{+}$ is the distribution function after collision.
Then we can take a combination of distribution function as
\begin{subequations}
\begin{equation}\label{eq17:a}
        \begin{aligned}
        f^{k}_{013}=&f^{k}_{013}-(s_0-s_2)(f^k_{478}-f^{k,eq}_{478})-s_0(f^k_{013}-f^{k,eq}_{013})-(s_0-s_2)(f^k_{256}-f^{k,eq}_{256}) \\
        &+[(\omega_1+2\omega_5)\theta (s_2-s_0)+(\omega_0+2\omega_1)(1-\frac{\theta s_0}{2})]\delta_tR, \\
        \end{aligned}
        \end{equation}
        \begin{equation}\label{eq17:b}
        f^{k+1}_{256}=f^{k}_{256}-(\frac{s_2}{2}-\frac{s_1}{2})(f^k_{478}-f^{k,eq}_{478})-(\frac{s_2}{2}+\frac{s_1}{2})(f^k_{256}-f^{k,eq}_{256})+(\omega_1+2\omega_5)(1-\frac{\theta s_2}{2})\delta_tR,
        \end{equation}
        \begin{equation}\label{eq17:c}
        f^{k-1}_{478}=f^{k}_{478}-(\frac{s_1}{2}+\frac{s_2}{2})(f^k_{478}-f^{k,eq}_{478})-(\frac{s_2}{2}-\frac{s_1}{2})(f^k_{256}-f^{k,eq}_{256})+(\omega_1+2\omega_5)(1-\frac{\theta s_2}{2})\delta_tR,
        \end{equation}
\end{subequations}
where $f^k_{ijm}=f^k_i+f^k_j+f^k_m$, $f^{k,eq}_{ijm}=f^{k,eq}_i+f^{k,eq}_j+f^{k,eq}_m$.
According to Eq. (\ref{eq12}), we can obtain
\begin{equation}\label{eq18}
f^k_{013}=\phi_k-f^k_{256}-f^k_{478}-\frac{\theta R}{2}\delta t.
\end{equation}
Substituting Eq. (\ref{eq18}) into Eq. (\ref{eq17:a}), one can obtain
\begin{equation}\label{eq19}
f^k_{256}+f^k_{478}=\phi_k-f^{k,eq}_{013}+A\delta_tR,
\end{equation}
where $a_0=\omega_0+2\omega_1$, $a_1=\omega_1+2\omega_5$, $A=-(a_0+a_1s_2\theta)/s_2$, with $\omega_1=\omega_2=\omega_3=\omega_4$,
$\omega_5=\omega_6=\omega_7=\omega_8$.
With the help of Eq. (\ref{eq19}), we can rewritten Eqs. (\ref{eq17:b}) and (\ref{eq17:c}) as
\begin{subequations}
        \begin{equation}\label{eq20:a}
        f^{k+1}_{256}=(1-s_1)f^k_{256}+s_1f_{256}^{k,eq}+B\delta_tR,
        \end{equation}
        \begin{equation}\label{eq20:b}
        f^{k-1}_{478}=(1-s_1)f^k_{478}+s_1f_{478}^{k,eq}+B\delta_tR,
        \end{equation}
\end{subequations}
where $B=a_1(1-{\theta s_2}/{2})-({s_2}-{s_1})A/2$.
Then we can get the following equation according to Eqs. (\ref{eq20:a}) and (\ref{eq20:b})
\begin{subequations}
        \begin{equation}\label{eq21:a}
        f^{k}_{256}=(1-s_1)f^{k-1}_{256}+s_1f_{256}^{k-1,eq}+B\delta_tR,
        \end{equation}
        \begin{equation}\label{eq21:b}
        f^{k}_{478}=(1-s_1)f^{k+1}_{478}+s_1f_{478}^{k+1,eq}+B\delta_tR.
        \end{equation}
\end{subequations}
With the help of Eq. (\ref{eq19}), Eqs. (\ref{eq21:a}) and (\ref{eq21:b}) can be written as
\begin{subequations}
        \begin{equation}\label{eq22:a}
        f^{k}_{256}=(1-s_1)(\phi_{k-1}-f^{k-1}_{478}-f^{k-1,eq}_{013}+A\delta_tR)+s_1f_{256}^{k-1,eq}+B\delta_tR,
        \end{equation}
        \begin{equation}\label{eq22:b}
        f^{k}_{478}=(1-s_1)(\phi_{k+1}-f^{k+1}_{256}-f^{k+1,eq}_{013}+A\delta_tR)+s_1f_{478}^{k+1,eq}+B\delta_tR.
        \end{equation}
\end{subequations}
Taking a sum of Eqs. (\ref{eq20:a}), (\ref{eq20:b}), (\ref{eq22:a}) and (\ref{eq22:b}), one can obtain
\begin{equation}\label{eq23}
a_1\frac{s_1-2}{s_1}(\phi_{k+1}+\phi_{k-1}-2\phi_k)=\frac{\phi_{k+1}u_{y,k+1}-\phi_{k-1}u_{y,k-1}}{2c}+\delta_tR,
\end{equation}
where Eq. (\ref{eq19}) has been adopted.
Then we can obtain the following equivalent difference equation of the MRT model,
\begin{equation}\label{eq24}
\frac{\phi_{k+1}u_{y,k+1}-\phi_{k-1}u_{y,k-1}}{2\delta x}=D\frac{\phi_{k+1}-2\phi_k+\phi_{k-1}}{\delta x^2}+R,
\end{equation}
where $D=({1}/{s_1}-{1}/{2})c^2_s\delta t$, $c^2_s=2a_1c^2$, $a_1=\omega_1+2\omega_5$.
Here we would like to point out that if we adopt different transform matrix $\textbf{M}$ which is constructed by orthogonal vectors, one can obtain the same equivalent difference equation \cite{zhao2019}.

Actually, for higher dimensions lattice velocity models (e.g., D3Q27), one can obtain the similar difference scheme as Eq. (\ref{eq24}) (see Appendix A for details).
Then we will get a useful equation, in the following derivation. When $k=1$, Eq. (\ref{eq21:b}) can be written as
\begin{equation}\label{eq25}
f^{1}_{478}=(1-s_1)f^{2}_{478}+s_1f_{478}^{2,eq}+B\delta_tR.
\end{equation}
Substituting Eq. (\ref{eq19}) into Eq. (\ref{eq25}), one can obtain
\begin{equation}\label{eq26}
f^{1}_{478}=(1-s_1)(\phi_2-f^{2,eq}_{013}+A\delta_tR-f^2_{256})+s_1f^{2,eq}_{478}+B\delta_tR.
\end{equation}
In addition, substituting Eq. (\ref{eq22:a}) into Eq. (\ref{eq26}) with the help of Eq. (\ref{eq3}) gives rise to
\begin{equation}\label{eq27}
f^{1}_{478}=(1-s_1)(2a_1\phi_2+A\delta_tR-(1-s_1)(2a_1\phi_1-f^1_{478}+A\delta_tR)+2a_1s_1\phi_1+B\delta_tR)+2a_1s_1\phi_2+B\delta_tR.
\end{equation}
We can rewrite the Eq. (\ref{eq27}) as
\begin{equation}\label{eq28}
s_1f^{1}_{478}=a_1\phi_2+(s_1-1)a_1\phi_1+\frac{(s_1A-B)(1-s_1)+B}{2-s_1}\delta_tR.
\end{equation}
\subsection{Discrete effect of the ABB boundary scheme}
\begin{figure}[ht]
 \centering
\includegraphics{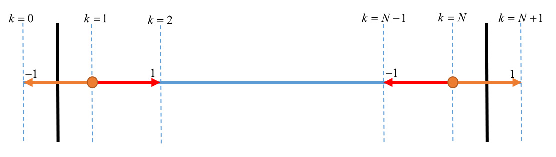}
\caption{The boundary arrangement in the D1Q3 lattice model; the black line denotes the boundary and is
located at $k = 1/2$ and $k = N + 1/2$.} \label{Fig2}
\end{figure}

To simplify the analysis on the discrete effect of the ABB boundary scheme,
a unidirectional and time-independent diffusion problem is adopted,
and it can be described by the following simplified equation and boundary conditions for one dimensional problem
\begin{equation}\label{eq29}
D\frac{\partial^2\phi}{\partial x^2}+R=0,
\end{equation}
\begin{equation}\label{eq30}
\phi(x=0)=\phi_0,\phi(x=L)=\phi_L,
\end{equation}
where $\phi_0$ and $\phi_L$ are constant, $L$ is the width and $D$ is the diffusion coefficient. $R$ is a constant source term, and is defined by
\begin{equation}\label{eq31}
R=2D\frac{\Delta\phi}{L^2},\Delta\phi=\phi_L-\phi_0.
\end{equation}
The analytical solution of the problem is given by
\begin{equation}\label{eq32}
\phi(x)=\phi_0+\frac{x}{L}(2-\frac{x}{L})\Delta\phi.
\end{equation}
Based on Eq. (\ref{eq15}), equivalent difference equation for the MRT model for Eq. (\ref{eq29}),
\begin{equation}\label{eq33}
D\frac{\phi_{k+1}-2\phi_k+\phi_{k-1}}{\delta x^2}+R=0.
\end{equation}
Then we can obtain the solution of Eq. (\ref{eq33}),
\begin{equation}\label{eq34}
\phi_k=-\frac{\Delta \phi}{N^2}k^2+ak+b,
\end{equation}
where $a$, $b$ are parameters to be determined.
If we consider ABB scheme, the value of $\phi$ at bottom and top boundaries can be given by
\begin{equation}\label{eq35}
\phi_{0.5}=\phi_0+\phi_s^{0.5},\quad \phi_{N+0.5}=\phi_L+\phi_s^{N+0.5}.
\end{equation}
where $\phi_s^{0.5}$, $\phi_s^{N+0.5}$ are numerical slip caused by ABB scheme, $N$ representing grid number.
Substituting Eq. (\ref{eq35}) into Eq. (\ref{eq34}), we obtain the numerical solution
\begin{equation}\label{eq36}
\phi_k=-\frac{\Delta \phi}{N^2}k^2+(2N+1)\frac{\Delta \phi}{N^2}k-(4N+1)\frac{\Delta \phi}{4N^2}+(k-\frac{1}{2})\frac{\phi_s^{N+0.5}-\phi_s^{0.5}}{N}+\phi_0+\phi_s^{0.5},
\end{equation}
In the following, we will focus on how to determine $\phi_s^{0.5}$ and $\phi_s^{N+0.5}$ from the ABB scheme.
As Fig. \ref{Fig2} shown,
the unknown distribution functions at the layers $k = 1$, $k = N$ can be determined by the following equations \cite{ZhangT},
\begin{equation}\label{eq37}
f^1_{1}=-f^{1,+}_{-1}+2\omega_1\phi_0,
\end{equation}
\begin{equation}\label{eq38}
f^N_{-1}=-f^{N,+}_{1}+2\omega_1\phi_L,
\end{equation}
where $f^{1,+}_{-1}$, $-f^{N,+}_{1}$ represent the distribution function after collision at the layers $k=1$ and $k=N$ respectively.
Following the process in Appendix B, we can get the numerical slip,
\begin{equation}\label{eq39}
\phi_s^{0.5}=\frac{4(2-s_1)\omega_0+s_2[-4+s_1+4(2-s_1)\omega_1\theta]}{4s_1s_2}\frac{\Delta \phi}{N^2},
\end{equation}
\begin{equation}\label{eq40}
\phi_s^{N+0.5}=\frac{4(2-s_1)\omega_0+s_2[-4+s_1+4(2-s_1)\omega_1\theta]}{4s_1s_2}\frac{\Delta \phi}{N^2}.
\end{equation}
As we can see, $\phi_s^{0.5}$ and $\phi_s^{N+0.5}$ have the same expression, thus we denote them by $\phi_s$ in the following discussion.
If the free parameter $s_2$ is chosen to satisfy the relation,
\begin{equation}\label{eq41}
4(2-s_1)\omega_0+s_2[-4+s_1+4(2-s_1)\omega_1\theta]=0,
\end{equation}
the discrete effect of the ABB scheme can be eliminated.

Furthermore, when we use the BGK model ($s_1=s_2$) to deal with the problem,
and take the weight coefficients $\omega_0$ and $\omega_1$  to satisfy Eq. (\ref{eq41}), the discrete effect on the ABB boundary scheme can also be eliminated.
However, this selection of the weight coefficients in the BGK model is limited due to the fact that the weight coefficients should be greater than 0 and less than 1.

Similarly, for the two-dimensional unidirectional steady problem with a parabolic distribution in one direction, we analyze the discrete effect in D2Q9 MRT model. For the ABB scheme,
\begin{subequations}
        \begin{equation}\label{eq42:a}
        f^1_{2}=-f^{1,+}_{4}+2\omega_1\phi_0,
        \end{equation}
        \begin{equation}\label{eq42:b}
        f^1_{5}=-f^{1,+}_{7}+2\omega_5\phi_0,
        \end{equation}
        \begin{equation}\label{eq42:c}
        f^1_{6}=-f^{1,+}_{8}+2\omega_5\phi_0,
        \end{equation}
\end{subequations}
where the $f^{1,+}_i=f_i(x_k,y_1,t)^{+}$ represent the distribution function after the collision.
Taking a sum of Eqs. (\ref{eq42:a}), (\ref{eq42:b}), and (\ref{eq42:c}), we obtain
\begin{equation}\label{eq43}
f^1_{256}=-f^{1,+}_{478}+2a_1\phi_0,
\end{equation}
which can be written as
\begin{equation}\label{eq44}
2a_1\phi_1+A\delta tR=s_1f^{1}_{478}-a_1s_1\phi_1-B\delta tR+2a_1\phi_0,
\end{equation}
with the help of Eqs. (\ref{eq19}) and (\ref{eq20:b}).
Substituting
\begin{subequations}
        \begin{equation}\label{eq45:a}
        \phi_1=\phi_0+\phi_s+(2-\frac{1}{2N})\frac{\Delta\phi}{2N},
        \end{equation}
        \begin{equation}\label{eq45:b}
        \phi_2=\phi_0+\phi_s+(2-\frac{3}{2N})\frac{3\Delta\phi}{2N},
        \end{equation}
\end{subequations}
and Eq. (\ref{eq28}) into Eq. (\ref{eq44}), we can obtain
\begin{equation}\label{eq46}
\phi_s=\frac{2a_0\Delta\phi}{N^2}[(\frac{1}{s_1}-\frac{1}{2})(\frac{1}{s_2}-\frac{1-2a_1\theta}{2a_0})-\frac{1}{8a_0}],
\end{equation}
where $a_0=\omega_0+2\omega_1$, $a_1=\omega_1+2\omega_5$ in D2Q9 model.

\begin{figure}[ht]
\centering
\subfigure[]{ \label{fig:mini:subfig:a0}
\includegraphics[scale=0.8]{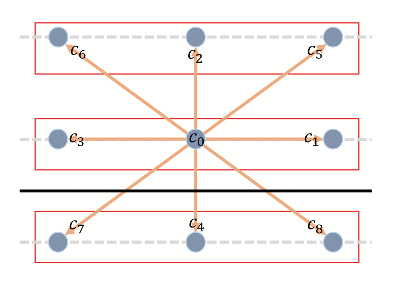}}
\subfigure[]{ \label{fig:mini:subfig:b0}
\includegraphics[scale=0.8]{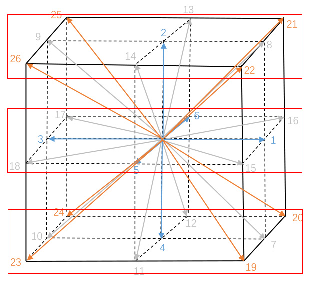}}
\caption{The Discrete velocity of D2Q9 and D3Q27, respectively.}
\label{Fig3}
\end{figure}

Similarly, for the three-dimensional unidirectional steady problem with a parabolic distribution in one direction,
one can obtain the following results with a similar derivation process,
\begin{equation}\label{eq47}
\phi_s=\frac{2a_0\Delta\phi}{N^2}[(\frac{1}{s_1}-\frac{1}{2})(\frac{1}{s_2}-\frac{1-2a_1\theta}{2a_0})-\frac{1}{8a_0}],
\end{equation}
where $c_s^2=2a_1c^2$, $a_0=\omega_0+4\omega_1+4\omega_7$, $a_1=\omega_1+4\omega_7+4\omega_{19}$ in D3Q19 model.
Taking the following equation
\begin{equation}\label{eq48}
\left(\frac{1}{s_2}-\frac{(a_0+2a_1(1-\theta))}{2a_0}\right)\left(\frac{1}{s_1}-\frac{1}{2}\right)=\frac{1}{8a_0},
\end{equation}
in Eq. (\ref{eq47}), one can eliminate the discrete effect.
The parameters $a_0$ and $a_1$ in the different lattice model are listed in Table \ref{Tab1_Model}, the velocities of D2Q9 and D3Q27 models are presented in Fig. {\ref{Fig3}}, and the relaxation factors $s_1$ and $s_2$ are associated with first and second moments.
We note that when $\theta=1$, $\omega_i=1/4(i=1-4)$ in D2Q4 model, $\omega_i=1/5(i=0-4)$ in D2Q5 model, $\omega_0=4/9, \omega_{1-4}=1/9,\omega_{5-8}=1/36$ in D2Q9 model, Eq. (\ref{eq47}) contains the previous works \cite{Cui}.
And Eq. (\ref{eq48}) is consist with the recent results \cite{Ginzburg2017} when $\theta=1$ in the frame of TRT model.
It should be noted that for a specified lattice model, we can determine the explicit expression of $\phi_s$ from Eq. (\ref{eq47}),
but the numerical slip $\phi_s$ could not be eliminated since $w_i$ is not flexible enough to satisfy Eq. (\ref{eq48}).
For example, in the D1Q2 model, ($\omega_0=0$, $\omega_1=1/2$), Eq. (\ref{eq48}) can not be satisfied under the condition of $0 < s_1 <2$ and $0< s_2 <2 $.
\begin{table}[ht]
\caption{The $a_0$ and $a_1$ in different lattice models.}
\centering
\begin{tabular}{cclcccccccc}
\hline \hline
&{Different models} & $\quad\quad$   &   $a_0$        & $\quad$  & $a_1$                 \\
\midrule[1pt]
&$D1Q2$      & & $0$       & & $\omega_1$    \\
&$D1Q3$      & & $\omega_0$       & & $\omega_1$    \\
&$D2Q4$      & & $2\omega_1$  & & $\omega_1$     \\
&$D2Q5$      & & $\omega_0+2\omega_1$  & & $\omega_1$     \\
&$D2Q9$      & & $\omega_0+2\omega_1$  & & $\omega_1+2\omega_5$     \\
&$D3Q7$      & & $\omega_0+4\omega_1$  & & $\omega_1$     \\
&$D3Q13$     & & $\omega_0+4\omega_1$  & & $4\omega_1$     \\
&$D3Q15$     & & $\omega_0+4\omega_1$  & & $\omega_1+4\omega_7$     \\
&$D3Q19$     & & $\omega_0+4\omega_1+4\omega_7$  & & $\omega_1+4\omega_7$     \\
&$D3Q27$     & & $\omega_0+4\omega_1+4\omega_7$  & & $\omega_1+4\omega_7+4\omega_{19}$     \\
 \hline \hline
\end{tabular}
\label{Tab1_Model}
\end{table}
\subsection{Discrete effect of the BB boundary scheme}
In this section, we analyze the BB boundary scheme under the same assumptions.
For D2Q9 model with the BB boundary scheme \cite{ladd1994numerical},
\begin{subequations}
        \begin{equation}\label{eq49:a}
        f^0_{2}=f^{0}_{4},
        \end{equation}
        \begin{equation}\label{eq49:b}
        f^0_{5}=f^{0}_{7},
        \end{equation}
        \begin{equation}\label{eq49:c}
        f^0_{6}=f^{0}_{8}.
        \end{equation}
\end{subequations}
Summing Eqs. (\ref{eq49:a}), (\ref{eq49:b}), and (\ref{eq49:c}), one can obtain
\begin{equation}\label{eq50}
f^0_{256}=f^{0}_{478}.
\end{equation}
Then Eq. (\ref{eq17:b}) can be written as
\begin{equation}\label{eqa5}
f^{1}_{256}=f^{0}_{478}-(\frac{s_2}{2}-\frac{s_1}{2})(f^0_{478}-f^{k,eq}_{478})-(\frac{s_2}{2}+\frac{s_1}{2})(f^0_{478}-f^{k,eq}_{256})+a_1(1-\frac{\theta s_2}{2})\delta_tR,
\end{equation}
where $a_1=\omega_1+2\omega_5$.
One can obtain
\begin{equation}\label{eq51}
2a_1f^{1}_{478}+AR-f_{478}^1=(1-s_2)[(1-s_1)f_{478}^1+s_1a_1\phi_1+BR]+a_1s_2\phi_0+a_1(1-\frac{\theta s_2}{2})\delta_tR.
\end{equation}
with the help of Eqs. (\ref{eq19}) and (\ref{eq20:b}).
Substituting
\begin{subequations}
        \begin{equation}\label{eq52:a}
        \phi_1=\phi_0+\phi_s+(2-\frac{1}{N})\frac{\Delta\phi}{N},
        \end{equation}
        \begin{equation}\label{eq52:b}
        \phi_2=\phi_0+\phi_s+(2-\frac{2}{N})\frac{2\Delta\phi}{N},
        \end{equation}
\end{subequations}
and Eq. (\ref{eq28}) into Eq. (\ref{eq51}), we can obtain
\begin{equation}\label{eq53}
\phi_s=\frac{2(2-s_1-s_2)\Delta\phi}{s_1s_2N}.
\end{equation}
\subsection{Discrete effect of the NEE boundary scheme}
For the NEE scheme \cite{zhao2002non},
\begin{equation}\label{eq54}
f^0_{k}=f^{0,eq}_{k}+(f^1_k-f^{1,eq}_k),
\end{equation}
Based on Eq. (\ref{eq54}), we obtain
\begin{equation}\label{eq55}
f^0_{256}=f^{0,eq}_{256}+(f^1_{256}-f^{1,eq}_{256}).
\end{equation}
According to Eq. (\ref{eq21:a}), we have
\begin{equation}\label{eq56}
f^{1}_{256}=(1-s_1)f^{0}_{256}+s_1f_{256}^{0,eq}+B\delta_tR,
\end{equation}
which can be written as
\begin{equation}\label{eq57}
f^{1}_{256}=(1-s_1)(f^{0,eq}_{256}+(f^1_{256}-f^{1,eq}_{256}))+s_1f_{256}^{0,eq}+B\delta_tR,
\end{equation}
with the help of Eq. (\ref{eq55}).
Substituting Eq. (\ref{eq19}) into Eq. (\ref{eq57}), one can obtain
\begin{equation}\label{eq58}
2a_1\phi_1+A\delta tR=s_1f^1_{478}+(1-s_1)(a_1\phi_0+a_1\phi_1+A\delta tR)+a_1s_1\phi_0+B\delta_tR.
\end{equation}
Substituting
\begin{subequations}
        \begin{equation}\label{eq59:a}
        \phi_1=\phi_0+\phi_s+(2-\frac{1}{N})\frac{\Delta\phi}{N},
        \end{equation}
        \begin{equation}\label{eq59:b}
        \phi_2=\phi_0+\phi_s+(2-\frac{2}{N})\frac{2\Delta\phi}{N},
        \end{equation}
\end{subequations}
and Eq. (\ref{eq28}) into Eq. (\ref{eq58}), one can obtain
\begin{equation}\label{eq60}
\phi_s=\frac{2(1-s_1)\Delta\phi}{s_1N^2}.
\end{equation}

\section{NUMERICAL RESULTS }
In this section, some simulations of CDEs are performed to test above analysis, and ABB scheme is employed to treat the Dirichlet boundary conditions.
In our simulations, the global relative error (GRE) and maximum error($E_{max}$) are used to measure accuracy, and are defined as
\begin{equation}\label{eq61}
\textrm{GRE} = \frac{\sqrt{\sum\limits_i |\phi(\textbf{x}_i,t) -\phi^{*}{(\textbf{x}_i,t)}|^2}}{\sqrt{\sum\limits_i|\phi^{*}{(\textbf{x}_i,t)}|^2}},\quad\textrm{E}_{max}=\mathop{max}\limits_{i}\{|\phi(\textbf{x}_i,t) -\phi^{*}{(\textbf{x}_i,t)}|\}
\end{equation}
where $\phi$ and $\phi^{*}$ are the numerical and analytical
solutions, respectively.
In addition, the following convergent criterion for the steady
problems is used,
\begin{equation}\label{eq62}
\frac{\sqrt{\sum\limits_i |\phi(\textbf{x}_i,t+1) -
\phi{(\textbf{x}_i,t)}|^2}}{\sqrt{\sum\limits_i
|\phi{(\textbf{x}_i,t)}|^2}}<10^{-9}.
\end{equation}
In our simulations, $f_i^{eq}$ is applied to approximate the initial distribution function $f_i$.
\subsection{Some unidirectional time-independent real-valued CDEs}
\subsubsection{A linear time-independent diffusion equation}
We first consider a two-dimensional linear time-independent diffusion equation with a constant source term,
\begin{equation}\label{eq63}
\begin{aligned}
&D\frac{\partial^2\phi}{\partial y^2}+R=0,\\
&\phi(x,y=0) = \phi_0,\quad \phi(x,y=L) =\phi_L,
\end{aligned}
\end{equation}
where $\phi_0$ and $\phi_L$ are two constants, $L$ is the width between
the top and bottom boundaries, and $R$ is the source term and is
defined by
\begin{equation}\label{eqa3}
R=\frac{2D\Delta \phi}{L^2}, \Delta \phi=\phi_L-\phi_0.
\end{equation}
The analytical solution of this problem is given by
\begin{equation}\label{eq64}
\phi(x,y)=\phi_0+\frac{y}{L}(2-\frac{y}{L})\Delta \phi.
\end{equation}
Here we consider the popular D2Q9 MRT model with $\theta =1$, the physical parameter $L = 1.0$, $u_x = 0.1$, $u_y = 0.0$, the diffusion coefficient $D = 0.1$, the boundary conditions $\phi_0=0$, $\phi_L=1$, $\delta_x=L/N$ with the grid number $N$ varying from 5 to 17.

First, we would like to verify that the parameters except $s_1$ and $s_2$ have little effect on numerical results.
In our simulations, the value of $s_1$ is determined by the diffusion coefficient, while $s_2$ is given by Eq. (\ref{eq48}).
We measured the GREs of the problem under different values of $s_3$, and present the results in Table \ref{Tab_Exp1_11} and Table \ref{Tab_Exp1_12}.
As shown in these table, for the fixed $s_1$ and $N$, the relaxation parameter $s_3$ has little influence on GREs.
For this reason, except $s_1$ and $s_2$, the other parameters in $\textbf{S}$ are set to be 1.0 in the following simulations. In general, the GRE decreases with the increase of grid number $N$, and as we shown in Table \ref{Tab_Exp1_11} the GRE increases for the accumulation of mechanical errors when the grid number $N$ increases.
\begin{table}[ht]
\caption{The GREs of D2Q9 MRT model with ABB boundary scheme and different relaxation parameters
 ($w_0={4}/{9}$, $w_1={1}/{9}$, $w_5={1}/{36}$).}
\centering
\begin{tabular}{cclcccccccc}
\hline \hline
\multicolumn{4}{c}{Different values}   $\quad$  & $N=5$             & $\quad$  & $N=9$            & $\quad$ & $N=17$                  \\
\midrule[1pt]
$s_1=0.1$       &$\quad$&  $s_3=0.0$&$\quad$& $1.6143\times10^{-14}$  & & $1.1575\times10^{-14}$  & &  $4.6266\times10^{-15}$    \\
                     &  &  $s_3=1.0$      & & $9.1778\times10^{-16}$  & & $4.5187\times10^{-16}$  & &  $3.2051\times10^{-16}$    \\
                     &  &  $s_3=s_1$      & & $6.4495\times10^{-16}$  & & $6.5046\times10^{-16}$  & &  $2.8975\times10^{-16}$    \\
                     &  &  $s_3=s_2$      & & $7.0977\times10^{-16}$  & & $5.5918\times10^{-16}$  & &  $6.5757\times10^{-16}$    \\
\midrule[0.5pt]
$s_1=0.6$            &  &  $s_3=0.0$      & & $1.4288\times10^{-14}$  & & $9.2039\times10^{-15}$  & &  $2.1330\times10^{-8}$    \\
                     &  &  $s_3=1.0$      & & $4.8550\times10^{-16}$  & & $2.4793\times10^{-15}$  & &  $2.1372\times10^{-8}$    \\
                     &  &  $s_3=s_1$      & & $2.6330\times10^{-16}$  & & $1.5328\times10^{-15}$  & &  $2.1355\times10^{-8}$    \\
                     &  &  $s_3=s_2$      & & $4.5732\times10^{-16}$  & & $4.3549\times10^{-15}$  & &  $2.1393\times10^{-8}$                      \\
\midrule[0.5pt]
$s_1=1.071797$       &  &  $s_3=0.0$      & & $2.1428\times10^{-14}$  & & $1.8222\times10^{-8}$  & &  $1.1939\times10^{-7}$    \\
                     &  &  $s_3=1.0$      & & $2.5713\times10^{-15}$  & & $1.8272\times10^{-8}$  & &  $1.1947\times10^{-7}$    \\
                     &  &  $s_3=s_1$      & & $2.3383\times10^{-15}$  & & $1.8275\times10^{-8}$  & &  $1.1948\times10^{-7}$    \\
                     &  &  $s_3=s_2$      & & $2.2578\times10^{-15}$  & & $1.8275\times10^{-8}$  & &  $1.1948\times10^{-7}$                      \\
\midrule[0.5pt]
$s_1=1.9$            &  &  $s_3=0.0$      & & $2.2912\times10^{-7}$  & & $8.1846\times10^{-7}$  & &  $3.0786\times10^{-6}$    \\
                     &  &  $s_3=1.0$      & & $2.2926\times10^{-7}$  & & $8.1861\times10^{-7}$  & &  $3.0787\times10^{-6}$    \\
                     &  &  $s_3=s_1$      & & $2.2938\times10^{-7}$  & & $8.1873\times10^{-7}$  & &  $3.0789\times10^{-6}$    \\
                     &  &  $s_3=s_2$      & & $2.2914\times10^{-7}$  & & $8.1849\times10^{-7}$  & &  $3.0786\times10^{-6}$                      \\
 \hline \hline
\end{tabular}
\label{Tab_Exp1_11}
\end{table}

\begin{table}[ht]
\caption{The $E_{max}$ of D2Q9 MRT model with ABB boundary scheme and different relaxation parameters
 ($w_0={4}/{9}$, $w_1={1}/{9}$, $w_5={1}/{36}$).}
\centering
\begin{tabular}{cclcccccccc}
\hline \hline
\multicolumn{4}{c}{Different values}   $\quad$  & $N=5$             & $\quad$  & $N=9$            & $\quad$ & $N=17$                  \\
\midrule[1pt]
$s_1=0.1$       &$\quad$&  $s_3=0.0$&$\quad$& $2.1427\times10^{-14}$  & & $1.5488\times10^{-14}$  & &  $4.6266\times10^{-15}$    \\
                     &  &  $s_3=1.0$      & & $1.3322\times10^{-16}$  & & $5.6899\times10^{-16}$  & &  $4.4409\times10^{-16}$    \\
                     &  &  $s_3=s_1$      & & $7.7716\times10^{-16}$  & & $7.7716\times10^{-16}$  & &  $4.4409\times10^{-16}$    \\
                     &  &  $s_3=s_2$      & & $7.7716\times10^{-16}$  & & $7.7716\times10^{-16}$  & &  $4.4409\times10^{-16}$    \\
\midrule[0.5pt]
$s_1=0.6$            &  &  $s_3=0.0$      & & $1.7097\times10^{-14}$  & & $1.0214\times10^{-15}$  & &  $2.2029\times10^{-8}$    \\
                     &  &  $s_3=1.0$      & & $4.4409\times10^{-16}$  & & $2.5535\times10^{-15}$  & &  $2.2073\times10^{-8}$    \\
                     &  &  $s_3=s_1$      & & $3.3307\times10^{-16}$  & & $1.6653\times10^{-15}$  & &  $2.2055\times10^{-8}$    \\
                     &  &  $s_3=s_2$      & & $4.4409\times10^{-16}$  & & $4.6629\times10^{-15}$  & &  $2.2095\times10^{-8}$                      \\
\midrule[0.5pt]
$s_1=1.071797$       &  &  $s_3=0.0$      & & $2.2759\times10^{-14}$  & & $1.8820\times10^{-8}$  & &  $1.2330\times10^{-7}$    \\
                     &  &  $s_3=1.0$      & & $2.4425\times10^{-15}$  & & $1.8871\times10^{-8}$  & &  $1.2339\times10^{-7}$    \\
                     &  &  $s_3=s_1$      & & $2.2204\times10^{-15}$  & & $1.8875\times10^{-8}$  & &  $1.2339\times10^{-7}$    \\
                     &  &  $s_3=s_2$      & & $2.2204\times10^{-15}$  & & $1.8875\times10^{-8}$  & &  $1.2339\times10^{-7}$                     \\
\midrule[0.5pt]
$s_1=1.9$            &  &  $s_3=0.0$      & & $2.3665\times10^{-7}$  & & $8.4530\times10^{-7}$  & &  $3.1796\times10^{-6}$    \\
                     &  &  $s_3=1.0$      & & $2.3679\times10^{-7}$  & & $8.4546\times10^{-7}$  & &  $3.1797\times10^{-6}$    \\
                     &  &  $s_3=s_1$      & & $2.3691\times10^{-7}$  & & $8.4558\times10^{-7}$  & &  $3.1798\times10^{-6}$    \\
                     &  &  $s_3=s_2$      & & $2.3667\times10^{-7}$  & & $8.4534\times10^{-7}$  & &  $3.1796\times10^{-6}$                      \\
 \hline \hline
\end{tabular}
\label{Tab_Exp1_12}
\end{table}

\begin{figure}[ht]
\centering
\subfigure[]{ \label{fig:mini:subfig:a1}
\includegraphics[scale=0.4]{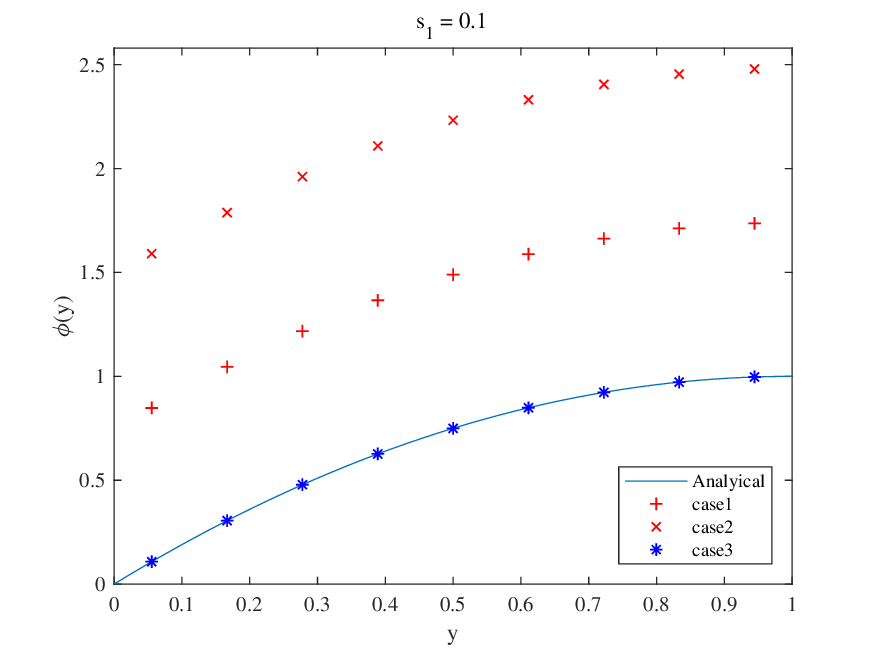}}
\subfigure[]{ \label{fig:mini:subfig:b1}
\includegraphics[scale=0.4]{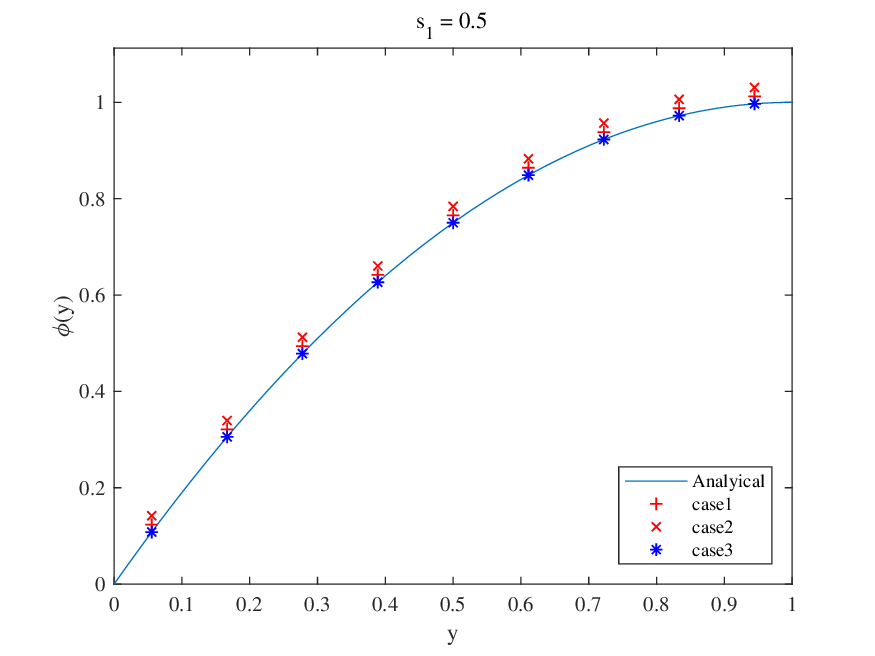}}
\caption{(Color online) D3Q19 BGK models with ABB boundary condition and the different weight coefficients.}
\label{Exp2_Fig1_1}
\end{figure}

After that, we test different weight coefficients in the D2Q9 BGK model when $s_1=0.1$ and $0.5$. In Fig. (\ref{Exp2_Fig1_1}), the case 1 is $\omega_i=1/9,(i=0-8)$, the case 2 is $\omega_0=4/9$, $\omega_1=1/9$, $\omega_5=1/36$, the case 3 is a set of weight coefficients satisfied Eq. (\ref{eq48}). In our simulation, case 3 is $\omega_0=1/1083$, $\omega_1=1/4332$, $\omega_5=1081/4332$ when $s_1=0.1$, $\omega_0=1/27$, $\omega_1=1/108$, $\omega_5=25/108$ when $s_1=0.5$. We can see that case 3 has more accurate results than case 1 and case 2.
As we known, the weight coefficients in the D2Q9 model are given as $\omega_0=4/9$, $\omega_1=1/9$, $\omega_5=1/36$ for Navier-Stokes equations.
Actually, weight coefficients in the LB model for CDEs are more flexible and they could be adjusted to give more accurate results.
This adjustment has certain limitations because the weight coefficients must be greater than zero. For BGK model, taking $s_1=s_2$, $0<a_0<1$, and $(1/s_1-1/2)^2=1/(8a_0)$, then we can get the limitation $1/s_1>(1+\sqrt{2})/2\sqrt{2}$. When $1/s_1>(1+\sqrt{2})/2\sqrt{2}$, $\phi_s$ on ABB boundary scheme can be eliminated with the adjustment of the weight coefficients in BGK model.
As for BB and NEE boundary schemes, when we consider BGK model(that is $s_1=s_2$), $\phi_s$ can be eliminated only if $s_1=1$. For this reason, the adjustment of ABB scheme is more flexible.

\begin{figure}[ht]
\centering
\subfigure[]{ \label{fig:mini:subfig:a2}
\includegraphics[scale=0.5]{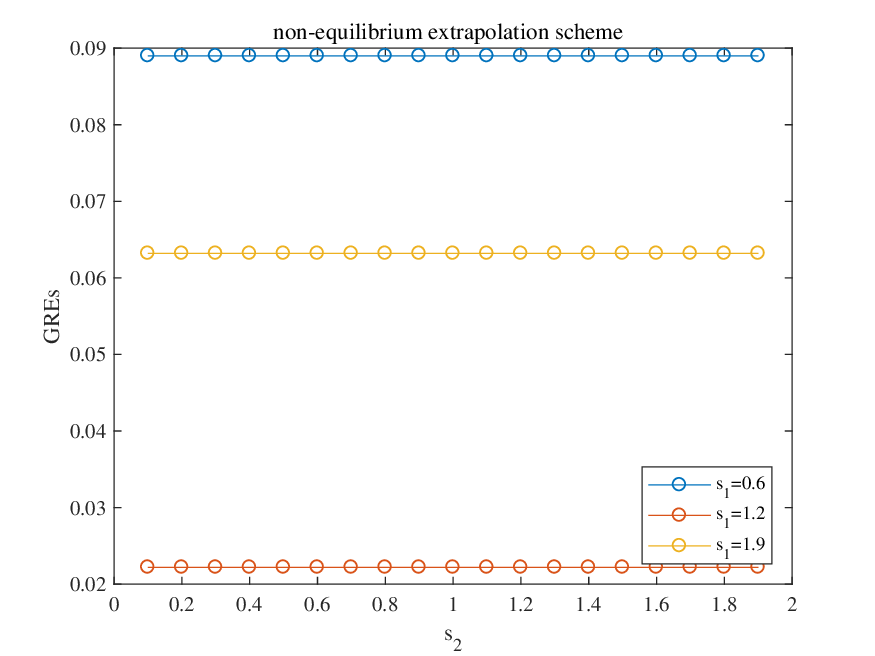}}
\caption{(Color online) D2Q9 MRT models with NEE boundary scheme.}
\label{Exp2_fneq1}
\end{figure}

$\phi_s$ of ABB boundary scheme depends on $s_1$, $s_2$, $\theta$ and $c_s^2$, $\phi_s$ of BB boundary scheme depends on $s_1$ and $s_2$, and $\phi_s$ of NEE scheme only depends on $s_1$.
We test the same problem with NEE scheme, taking $s_1=0.6, 1.2, 1.9$, with different $s_2$.
As we shown in Fig. (\ref{Exp2_fneq1}), $s_2$ has little effect on numerical results. And when we change the value of $s_1$,
we can see that the GRE has a minimum when $s_1=1$, which agree with Eq. (\ref{eq60}).
\begin{figure}[ht]
\centering
\subfigure[]{ \label{fig:mini:subfig:b2}
\includegraphics[scale=0.5]{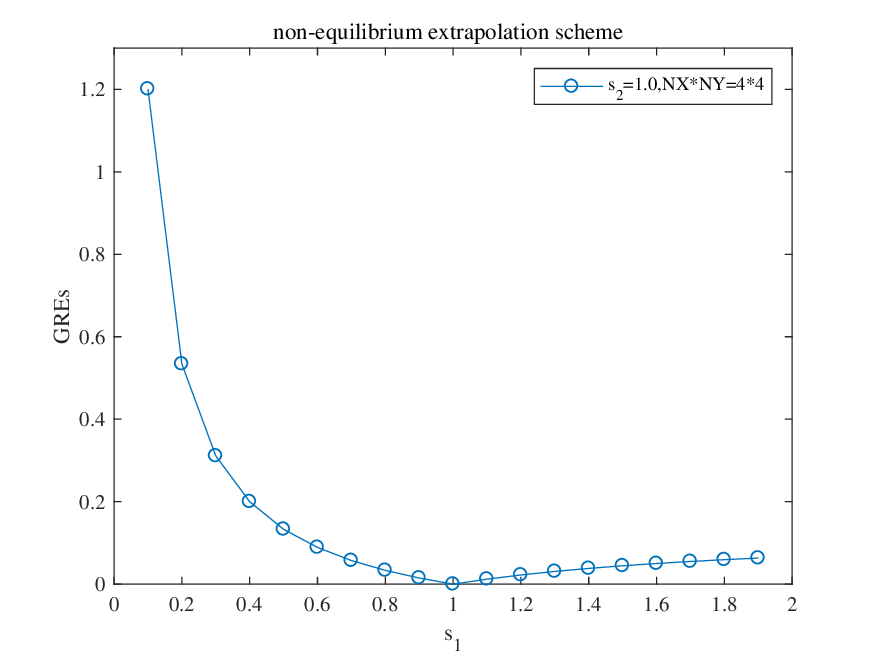}}
\caption{(Color online) D2Q9 MRT models with NEE boundary scheme.}
\label{Exp2_fneq2}
\end{figure}
Then we consider $\phi_s$ of BB scheme as Eq. (\ref{eq53}). When $s_2=2-s_1$, the discrete effect can be eliminated for
the unidirectional steady problem with a parabolic distribution in one direction.
We take a simulation of the same problem as Eq. (\ref{eq63}) with BB scheme, taking $s_1=0.1, 0.6, 1.0, 1.9$ respectively, and shown the result in Fig. (\ref{Exp2_BB}).
Under the same lattice size to eliminate the numerical slip in MRT model, we can adjust the parameter $s_2$ to satisfy $s_1+s_2=2$ for BB boundary scheme while in the BGK model $s_2$ is determined by diffusion coefficient, and can not be adjusted. As the figures shown, we can adjust $s_2$ to get more accurate results.
\begin{figure}[ht]
\centering
\subfigure[]{ \label{fig:mini:subfig:a3}
\includegraphics[scale=0.4]{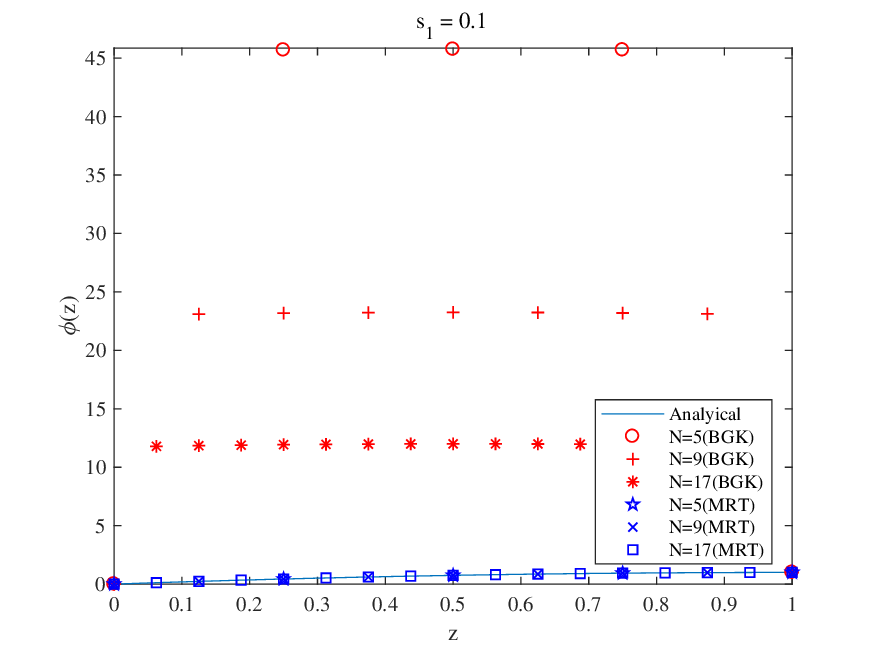}}
\subfigure[]{ \label{fig:mini:subfig:b3}
\includegraphics[scale=0.4]{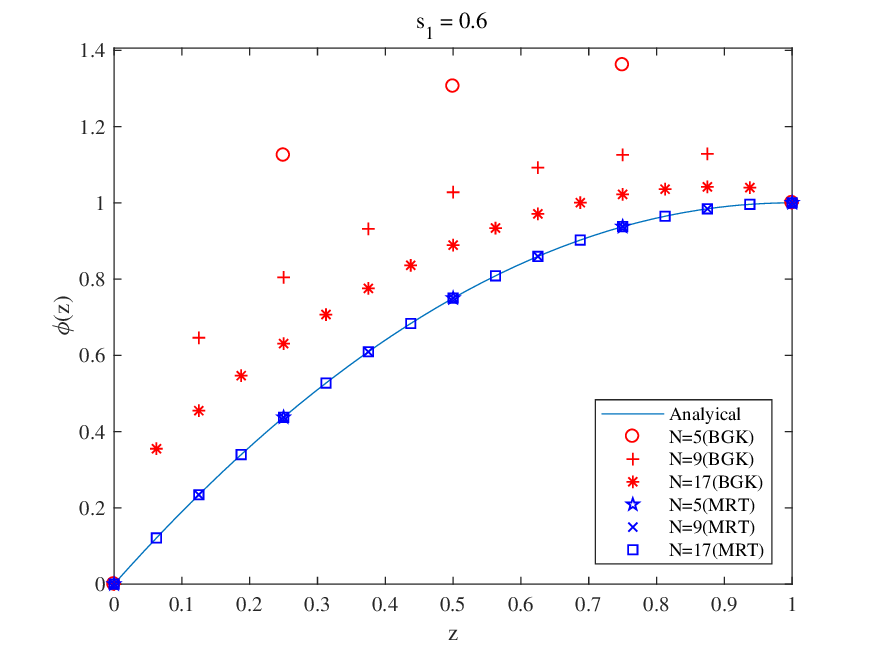}}
\subfigure[]{ \label{fig:mini:subfig:c3}
\includegraphics[scale=0.4]{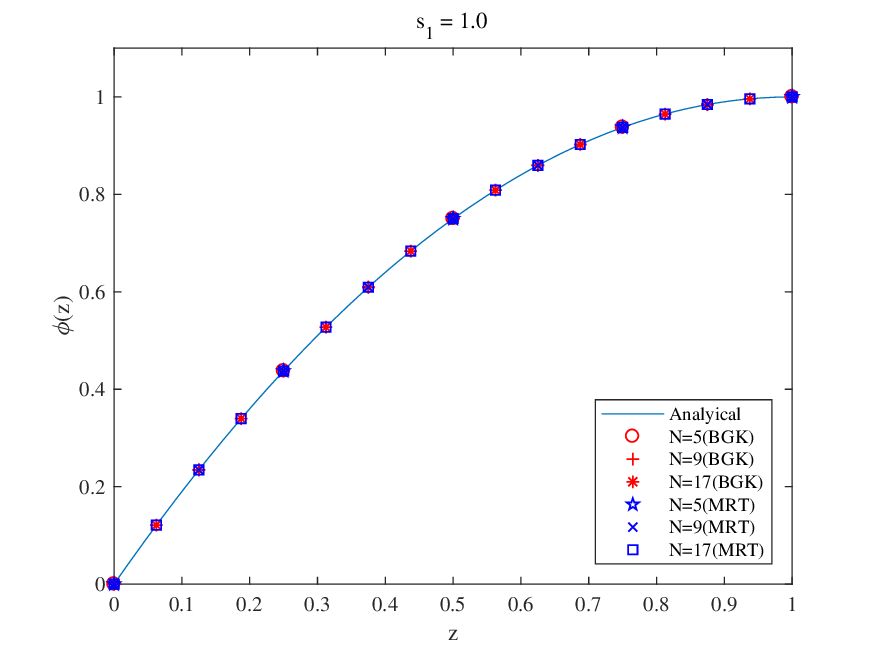}}
\subfigure[]{ \label{fig:mini:subfig:d3}
\includegraphics[scale=0.4]{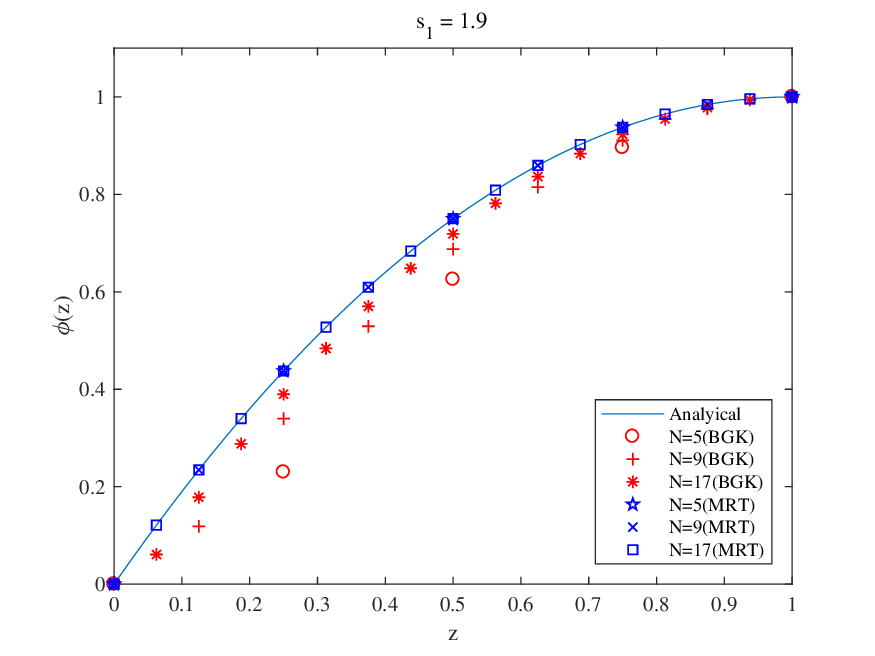}}
\caption{(Color online) D2Q9 BGK and MRT models with BB boundary scheme and the weight coefficients $\omega_0={4}/{9},\omega_1={1}/{9},\omega_7={1}/{36}$.}
\label{Exp2_BB}
\end{figure}

\begin{figure}[ht]
\centering
\subfigure[]{ \label{fig:mini:subfig:a4}
\includegraphics[scale=0.4]{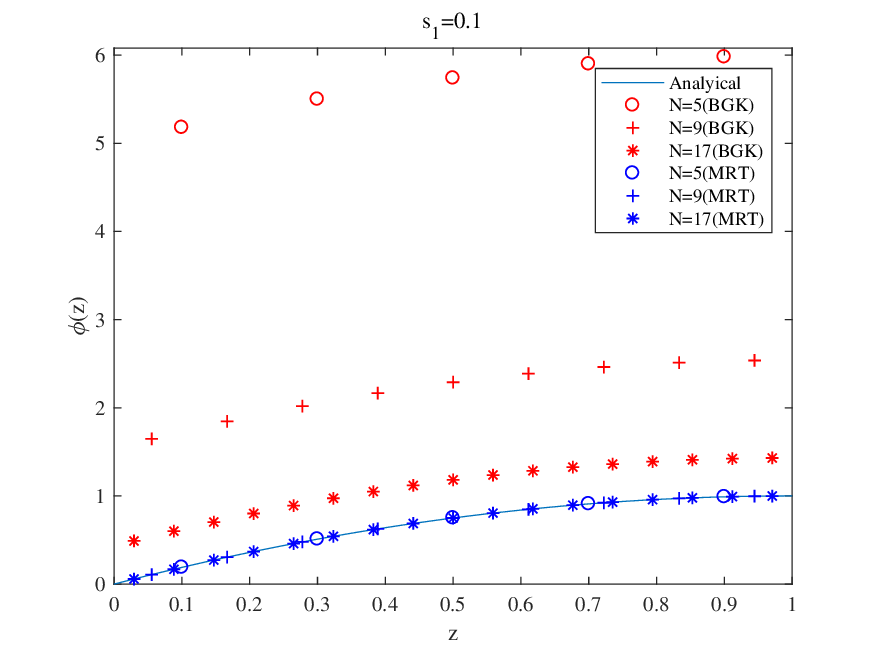}}
\subfigure[]{ \label{fig:mini:subfig:b4}
\includegraphics[scale=0.4]{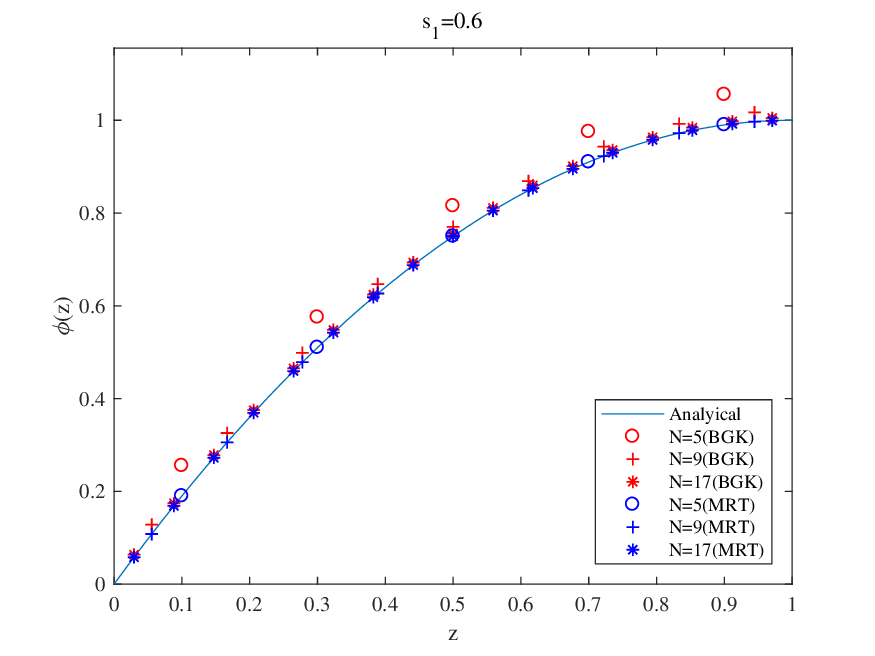}}
\subfigure[]{ \label{fig:mini:subfig:c4}
\includegraphics[scale=0.4]{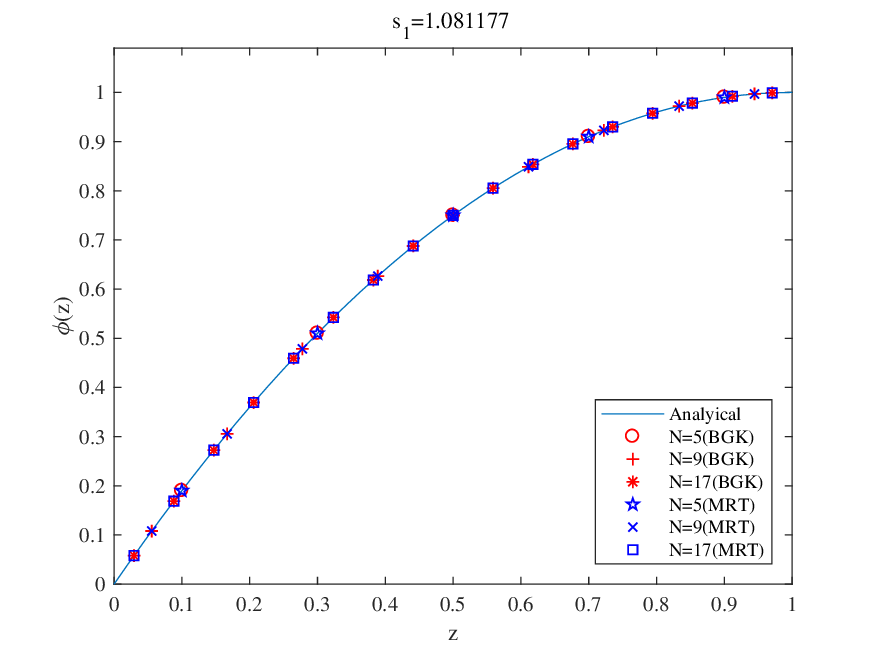}}
\subfigure[]{ \label{fig:mini:subfig:d4}
\includegraphics[scale=0.4]{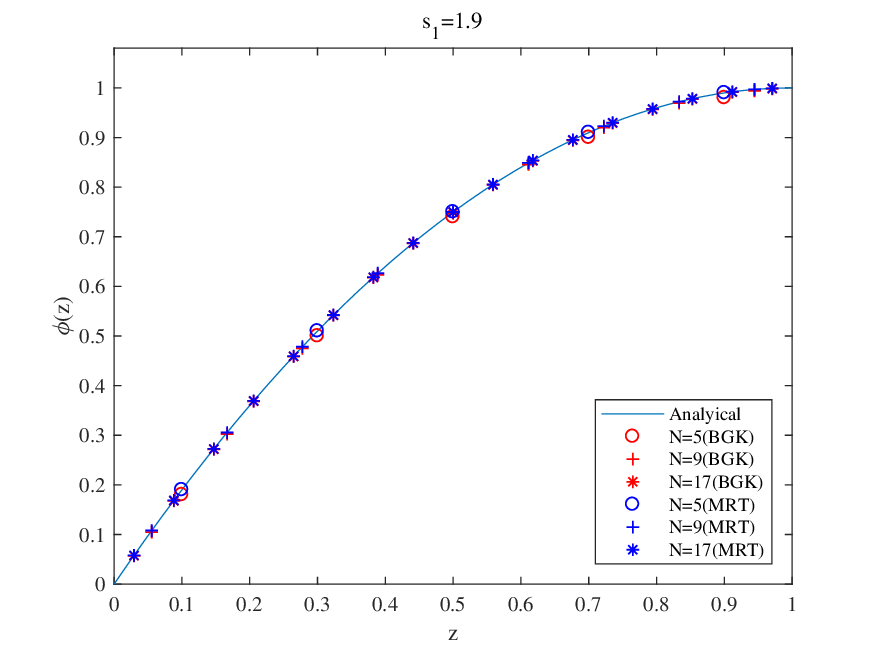}}
\caption{(Color online) D3Q19 BGK and MRT models with ABB boundary scheme and the weight coefficients $\omega_0={16}/{52},\omega_1={4}/{52},\omega_7={1}/{52}$.}
\label{Exp2_Fig1}
\end{figure}

\begin{figure}[ht]
\centering
\subfigure[]{ \label{fig:mini:subfig:a5}
\includegraphics[scale=0.4]{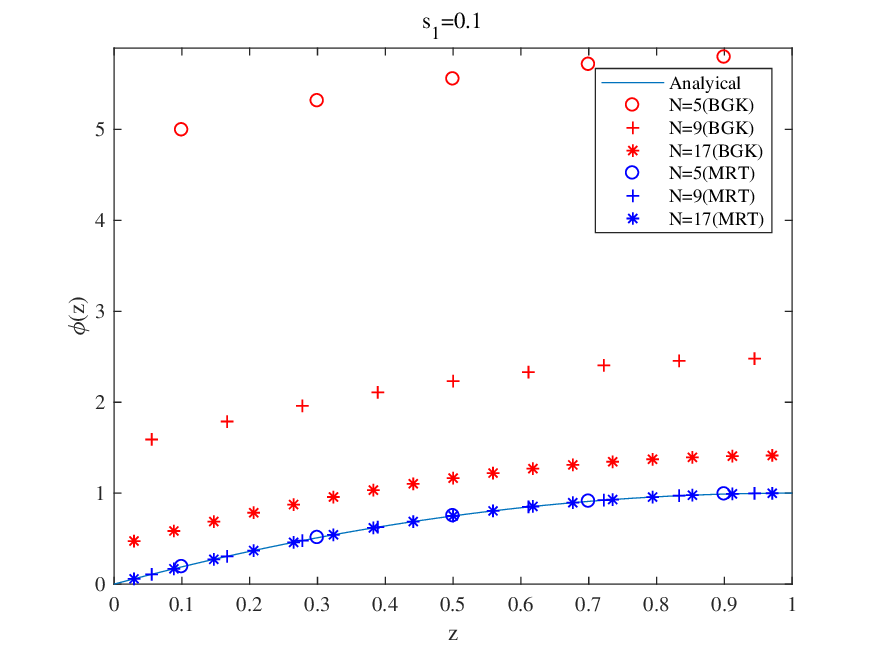}}
\subfigure[]{ \label{fig:mini:subfig:b5}
\includegraphics[scale=0.4]{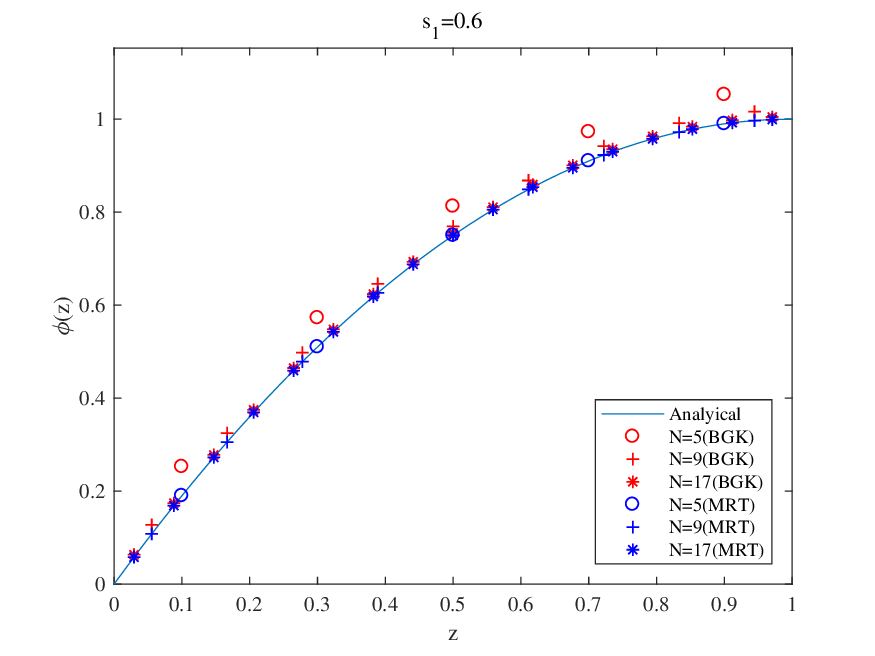}}
\subfigure[]{ \label{fig:mini:subfig:c5}
\includegraphics[scale=0.4]{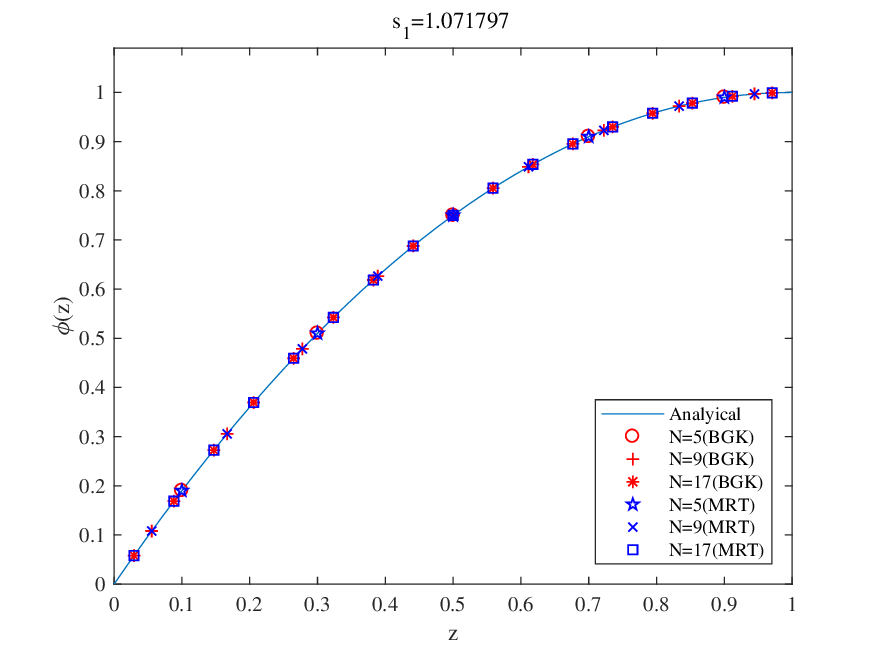}}
\subfigure[]{ \label{fig:mini:subfig:d5}
\includegraphics[scale=0.4]{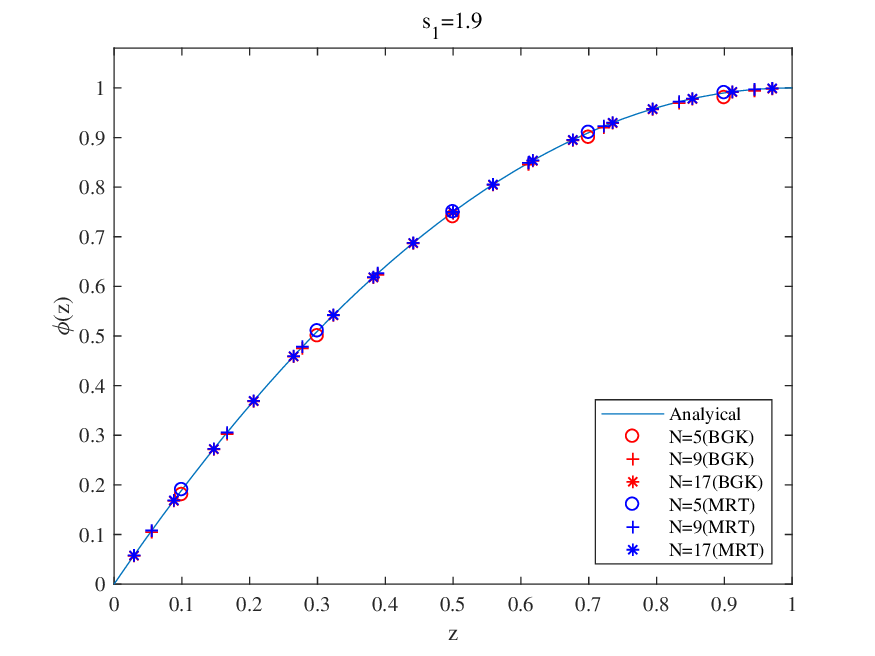}}
\caption{(Color online) D3Q19 BGK and MRT models with ABB boundary scheme and the weight coefficients $\omega_0={1}/{4},\omega_1={1}/{12},\omega_7={1}/{48}$).}
\label{Exp2_Fig2}
\end{figure}

\begin{figure}[ht]
\centering
\subfigure[]{ \label{fig:mini:subfig:a6}
\includegraphics[scale=0.4]{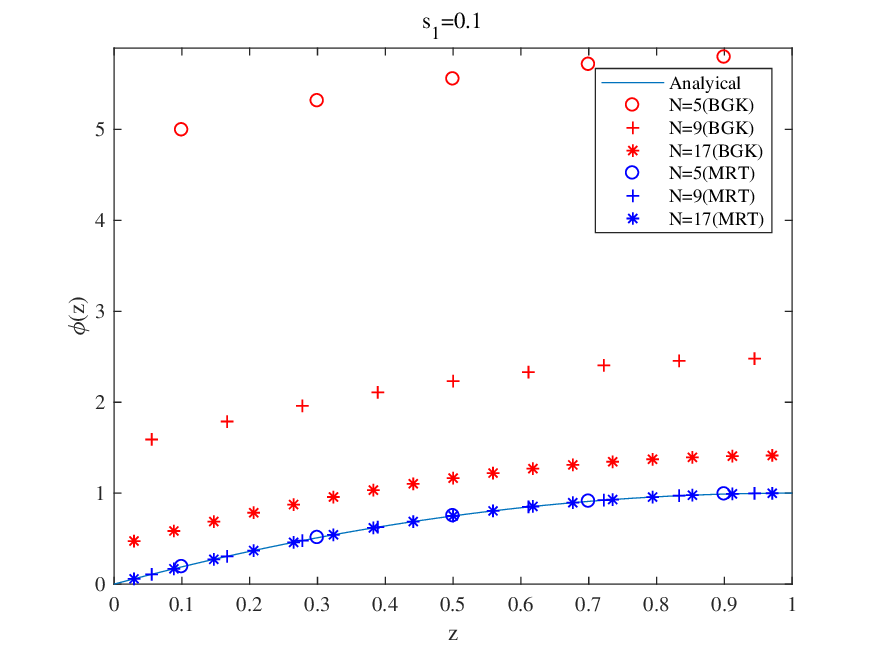}}
\subfigure[]{ \label{fig:mini:subfig:b6}
\includegraphics[scale=0.4]{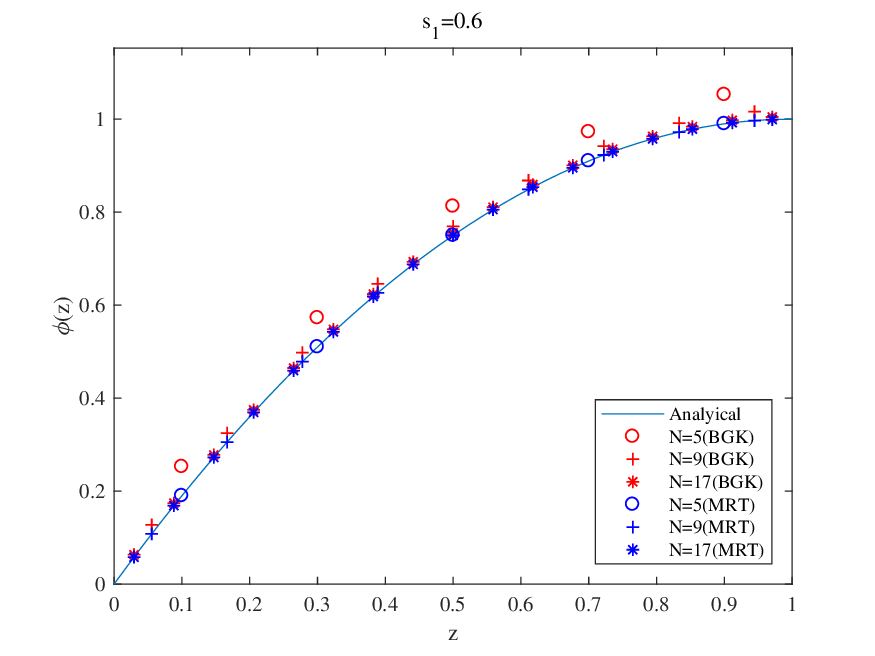}}
\subfigure[]{ \label{fig:mini:subfig:c6}
\includegraphics[scale=0.4]{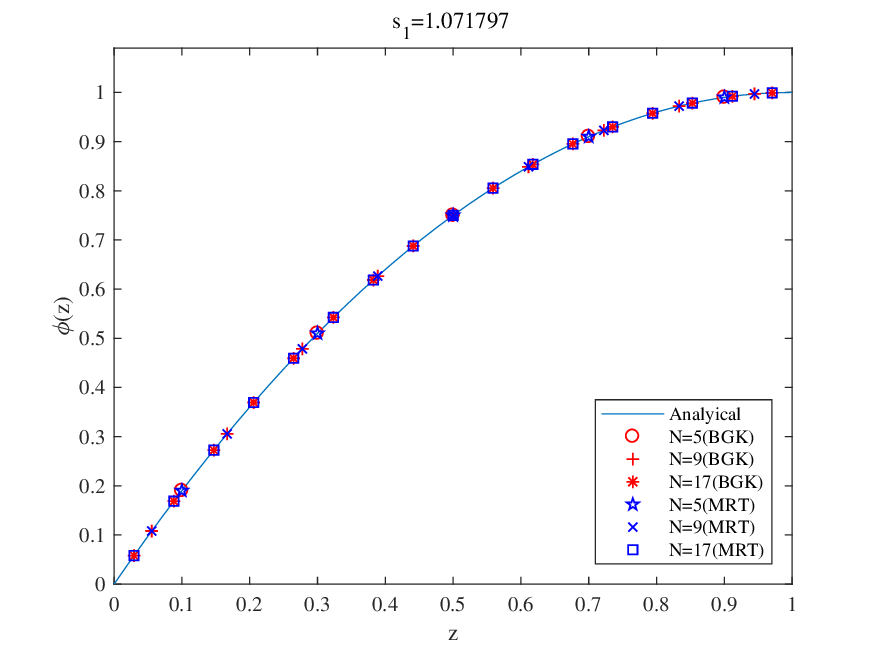}}
\subfigure[]{ \label{fig:mini:subfig:d6}
\includegraphics[scale=0.4]{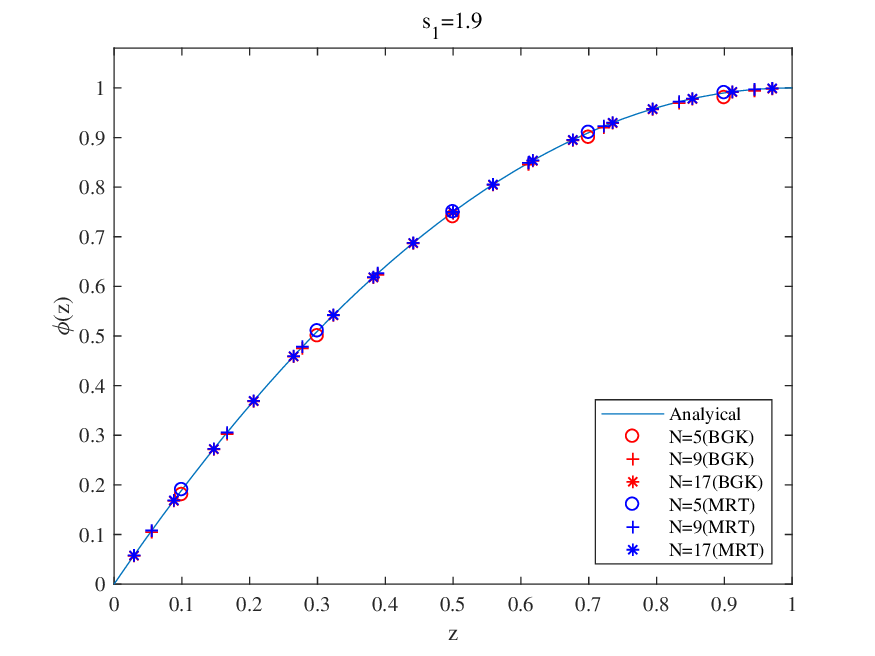}}
\caption{(Color online) D3Q19 BGK and MRT models with ABB boundary scheme and the weight coefficients $\omega_0={1}/{3},\omega_1={1}/{18},\omega_7={1}/{36}$.}
\label{Exp2_Fig3}
\end{figure}

\begin{figure}[ht]
\centering
\subfigure[]{ \label{fig:mini:subfig:a7}
\includegraphics[scale=0.4]{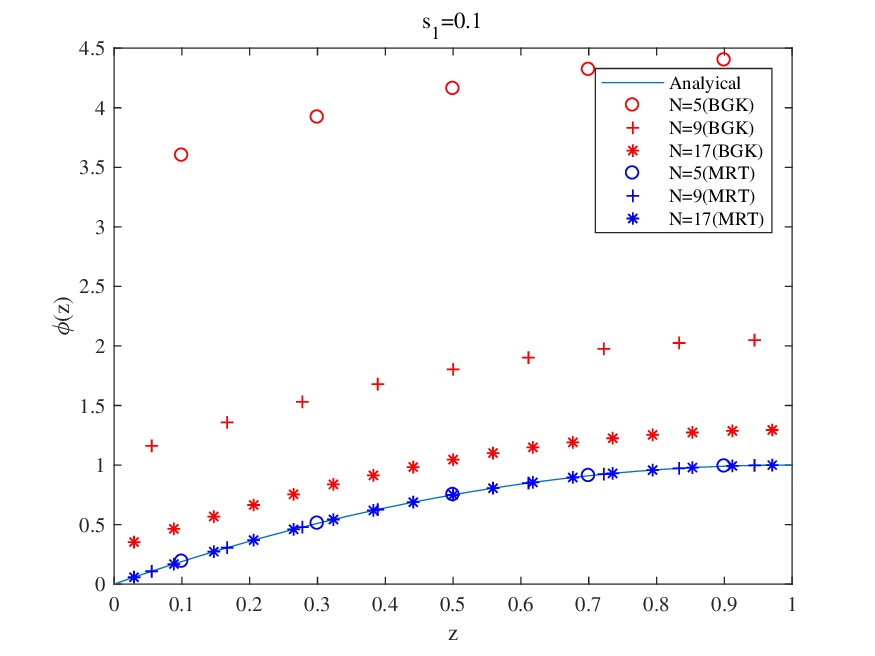}}
\subfigure[]{ \label{fig:mini:subfig:b7}
\includegraphics[scale=0.4]{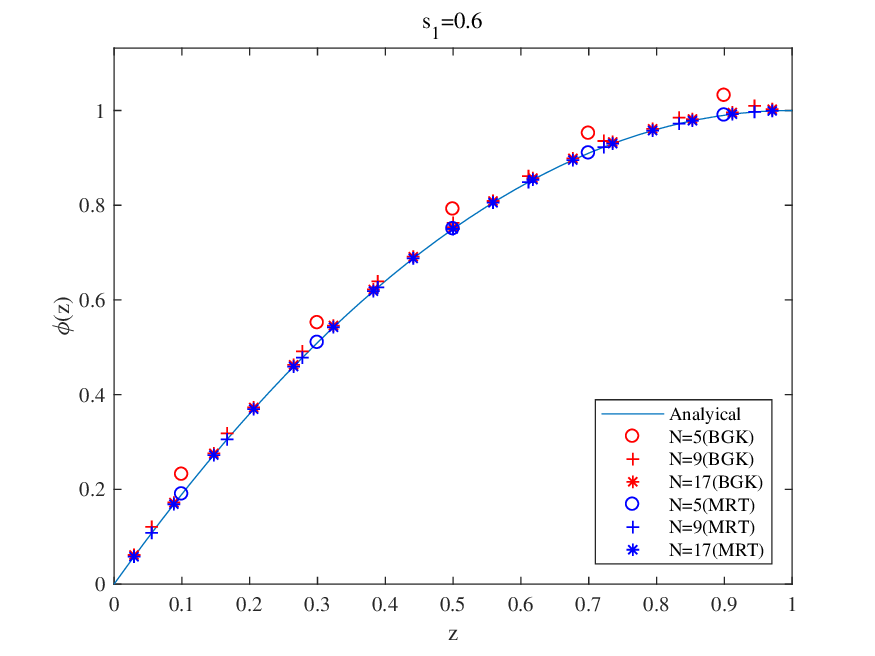}}
\subfigure[]{ \label{fig:mini:subfig:c7}
\includegraphics[scale=0.4]{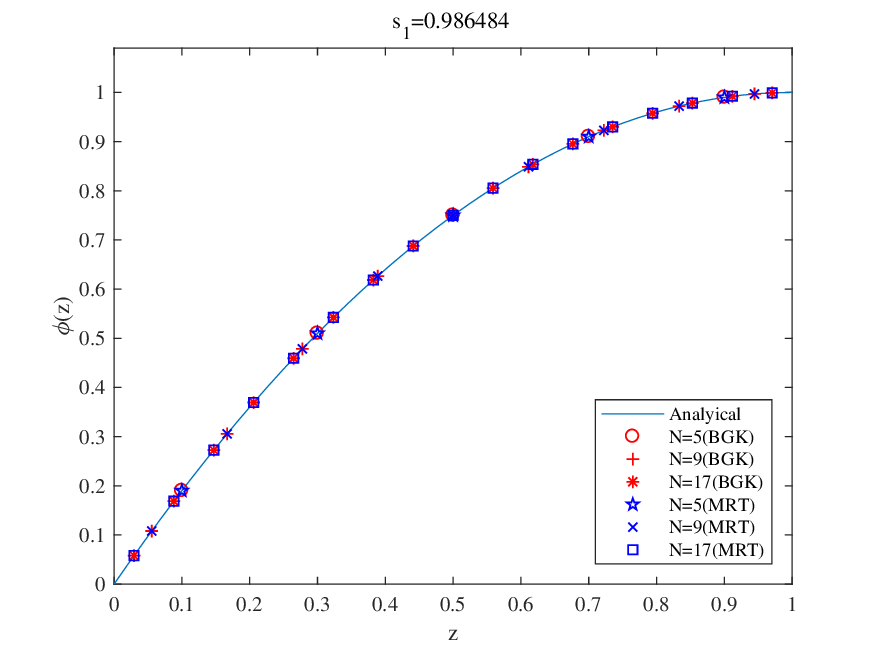}}
\subfigure[]{ \label{fig:mini:subfig:d7}
\includegraphics[scale=0.4]{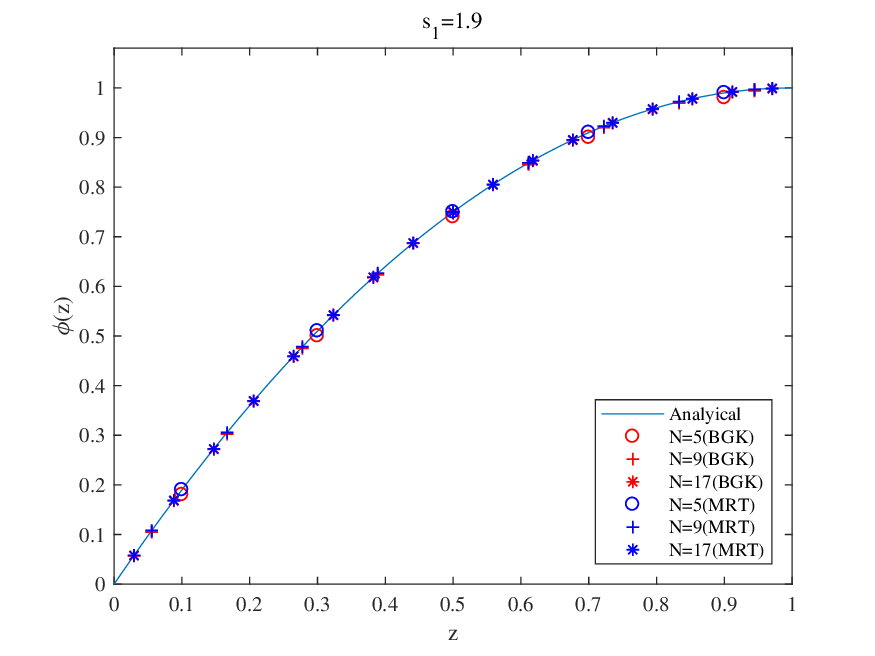}}
\caption{(Color online) D3Q19 BGK and MRT models with ABB boundary scheme and the weight coefficients
$\omega_0={1}/{19},(i=0-18)$.}
\label{Exp2_Fig4}
\end{figure}

Then, we consider a three-dimensional linear time-independent diffusion equation with a constant source term,
\begin{equation}\label{eqa1}
\begin{aligned}
&D\frac{\partial^2\phi}{\partial z^2}+R=0,\\
&\phi(x,y,z=0) = \phi_0,\quad \phi(x,y,z=L) =\phi_L.
\end{aligned}
\end{equation}
The analytical solution of this problem is given by
\begin{equation}\label{eqa2}
\phi(x,y,z)=\phi_0+\frac{z}{L}(2-\frac{z}{L})\Delta \phi.
\end{equation}
Here we consider the popular D3Q19 BGK and MRT model, the physical parameters $L = 1.0$, $u_x = 0.1$, $u_y = 0.0$, $u_z = 0.0$, the diffusion coefficient $D = 0.1$, the boundary conditions $\phi_0=0$, $\phi_L=1$, $\delta_x=L/N$ with the grid number $N$ varying from 5 to 17.
For the ABB boundary scheme, we can adjust the parameter $s_2$ to satisfy Eq. (\ref{eq48}) to get more accurate results.
We perform some simulations with both BGK and MRT models, and present the results in Figs. \ref{Exp2_Fig1}, \ref{Exp2_Fig2}, \ref{Exp2_Fig3} and \ref{Exp2_Fig4}.
In these figures, the values of $s_1$ are taken to be $0.1$, $0.6$, $1.9$, and a particular value satisfying Eq. (\ref{eq48}) under the condition of $s_1=s_2$.
From the results in Figs. \ref{Exp2_Fig1}, \ref{Exp2_Fig2}, \ref{Exp2_Fig3} and \ref{Exp2_Fig4}, one can see that when $s_2$ satisfies Eq. (\ref{eq48}), the numerical results are in good agreement with analytical solutions.

Here we give some comparisons of the GRE and $E_{max}$ among D2Q5 and D2D9, D3Q7 and D3Q19 models in Tables. \ref{Tab_Exp3_1}, \ref{Tab_Exp3_3}, \ref{Tab_Exp3_2} and \ref{Tab_Exp3_4},
and find that there are no apparent differences among D2Q5 and D2D9, D3Q7 and D3Q19 models when we adjust $s_2$ to satisfy Eq. (\ref{eq48}) for ABB boundary scheme.
However, the D2Q5 and D3Q7 models are more efficient since less discrete velocities are included.
\begin{table}[ht]
\caption{The GREs of D2Q5 and D2Q9 MRT models with ABB boundary scheme and different parameters.}
  \centering
\begin{tabular}{cclcccccccc}
\hline \hline
\multicolumn{4}{c}{Different models}   $\quad$  & $N=5$             & $\quad$  & $N=9$            & $\quad$ & $N=17$                  \\
\midrule[1pt]
$s_1=0.1$       &$\quad$&  $D2Q9,\omega_0=\frac{4}{9},\omega_1=\frac{1}{9},\omega_5=\frac{1}{36}$&$\quad$& $9.1778\times10^{-16}$  & & $4.5187\times10^{-16}$  & &  $3.2051\times10^{-16}$    \\
                   &  &  $D2Q5,\omega_0=\frac{1}{5},\omega_1=\frac{1}{5}$                         & & $5.7786\times10^{-16}$  & & $5.2053\times10^{-16}$  & &  $3.3281\times10^{-16}$    \\
\midrule[0.5pt]
$s_1=0.6$            &  &  $D2Q9,\omega_0=\frac{4}{9},\omega_1=\frac{1}{9},\omega_5=\frac{1}{36}$&$\quad$& $4.8550\times10^{-16}$  & & $2.4793\times10^{-15}$  & &  $2.1372\times10^{-8}$    \\
                     &  &  $D2Q5,\omega_0=\frac{1}{5},\omega_1=\frac{1}{5}$                         & & $2.8491\times10^{-16}$  & & $1.5632\times10^{-16}$  & &  $1.5599\times10^{-8}$    \\
\midrule[0.5pt]
$s_1=1.9$            &  &  $D2Q9,\omega_0=\frac{4}{9},\omega_1=\frac{1}{9},\omega_5=\frac{1}{36}$&$\quad$& $2.2926\times10^{-7}$  & & $8.1861\times10^{-7}$  & &  $3.0787\times10^{-6}$    \\
                     &  &  $D2Q5,\omega_0=\frac{1}{5},\omega_1=\frac{1}{5}$                         & & $1.4640\times10^{-7}$  & & $6.7983\times10^{-7}$  & &  $2.6060\times10^{-6}$    \\
 \hline \hline
\end{tabular}
\label{Tab_Exp3_1}
\end{table}

\begin{table}[ht]
\caption{The $E_{max}$ of D2Q5 and D2Q9 MRT models with ABB boundary scheme and different parameters.
  }\centering
\begin{tabular}{cclcccccccc}
\hline \hline
\multicolumn{4}{c}{Different models}   $\quad$  & $N=5$             & $\quad$  & $N=9$            & $\quad$ & $N=17$                  \\
\midrule[1pt]
$s_1=0.1$       &$\quad$&  $D2Q9,\omega_0=\frac{4}{9},\omega_1=\frac{1}{9},\omega_5=\frac{1}{36}$&$\quad$& $1.3322\times10^{-15}$  & & $5.6899\times10^{-16}$  & &  $4.4409\times10^{-16}$    \\
                   &  &  $D2Q5,\omega_0=\frac{1}{5},\omega_1=\frac{1}{5}$                         & & $6.1062\times10^{-16}$  & & $5.5511\times10^{-16}$  & &  $4.4409\times10^{-16}$    \\
\midrule[0.5pt]
$s_1=0.6$            &  &  $D2Q9,\omega_0=\frac{4}{9},\omega_1=\frac{1}{9},\omega_5=\frac{1}{36}$&$\quad$& $4.4409\times10^{-16}$  & & $2.5535\times10^{-15}$  & &  $2.2073\times10^{-8}$    \\
                     &  &  $D2Q5,\omega_0=\frac{1}{5},\omega_1=\frac{1}{5}$                         & & $3.3307\times10^{-16}$  & & $2.2204\times10^{-16}$  & &  $1.6110\times10^{-8}$    \\
\midrule[0.5pt]
$s_1=1.9$            &  &  $D2Q9,\omega_0=\frac{4}{9},\omega_1=\frac{1}{9},\omega_5=\frac{1}{36}$&$\quad$& $2.3679\times10^{-7}$  & & $8.4546\times10^{-7}$  & &  $3.1797\times10^{-6}$    \\
                     &  &  $D2Q5,\omega_0=\frac{1}{5},\omega_1=\frac{1}{5}$                         & & $1.5121\times10^{-7}$  & & $7.0213\times10^{-7}$  & &  $2.5966\times10^{-6}$    \\
 \hline \hline
\end{tabular}
\label{Tab_Exp3_3}
\end{table}

\begin{table}[ht]
\caption{The GREs of D3Q7 and D3Q19 MRT models with ABB boundary scheme and different parameters.}
  \centering
\begin{tabular}{cclcccccccc}
\hline \hline
\multicolumn{4}{c}{Different models}   $\quad$  & $N=5$             & $\quad$  & $N=9$            & $\quad$ & $N=17$                  \\
\midrule[1pt]
$s_1=0.1$       &$\quad$&  $D3Q19,\omega_0=\frac{1}{3},\omega_1=\frac{1}{18},\omega_7=\frac{1}{36}$&$\quad$& $3.0474\times10^{-10}$  & & $1.7407\times10^{-10}$  & &  $1.9787\times10^{-10}$    \\
                     &  &  $D3Q7,\omega_0=\frac{1}{3},\omega_1=\frac{1}{9}$                         & & $1.5758\times10^{-10}$  & & $7.4854\times10^{-11}$  & &  $5.6273\times10^{-11}$    \\

\midrule[0.5pt]
$s_1=0.6$            &  &  $D3Q19,\omega_0=\frac{1}{3},\omega_1=\frac{1}{18},\omega_7=\frac{1}{36}$&$\quad$& $3.2280\times10^{-9}$  & & $1.7372\times10^{-9}$  & &  $2.1372\times10^{-8}$    \\
                     &  &  $D3Q7,\omega_0=\frac{1}{3},\omega_1=\frac{1}{9}$                         & & $4.4101\times10^{-11}$  & & $2.4137\times10^{-9}$  & &  $8.7858\times10^{-8}$    \\
\midrule[0.5pt]
$s_1=1.9$            &  &  $D3Q19,\omega_0=\frac{1}{3},\omega_1=\frac{1}{18},\omega_7=\frac{1}{36}$&$\quad$& $2.2926\times10^{-7}$  & & $8.1861\times10^{-7}$  & &  $3.0787\times10^{-6}$    \\
                     &  &  $D3Q7,\omega_0=\frac{1}{3},\omega_1=\frac{1}{9}$                         & & $3.1045\times10^{-7}$  & & $1.2000\times10^{-6}$  & &  $4.6762\times10^{-6}$    \\
 \hline \hline
\end{tabular}
\label{Tab_Exp3_2}
\end{table}
\begin{table}[ht]
\caption{The $E_{max}$ of D3Q7 and D3Q19 MRT models with ABB boundary scheme and different parameters.}
  \centering
\begin{tabular}{cclcccccccc}
\hline \hline
\multicolumn{4}{c}{Different models}   $\quad$  & $N=5$             & $\quad$  & $N=9$            & $\quad$ & $N=17$                  \\
\midrule[1pt]
$s_1=0.1$       &$\quad$&  $D3Q19,\omega_0=\frac{1}{3},\omega_1=\frac{1}{18},\omega_7=\frac{1}{36}$&$\quad$& $3.7204\times10^{-10}$  & & $1.7549\times10^{-10}$  & &  $3.1573\times10^{-10}$    \\
                     &  &  $D3Q7,\omega_0=\frac{1}{3},\omega_1=\frac{1}{9}$                         & & $1.2115\times10^{-10}$  & & $1.1272\times10^{-11}$  & &  $6.4008\times10^{-11}$    \\

\midrule[0.5pt]
$s_1=0.6$            &  &  $D3Q19,\omega_0=\frac{1}{3},\omega_1=\frac{1}{18},\omega_7=\frac{1}{36}$&$\quad$& $3.3340\times10^{-9}$  & & $1.7942\times10^{-9}$  & &  $2.2073\times10^{-8}$    \\
                     &  &  $D3Q7,\omega_0=\frac{1}{3},\omega_1=\frac{1}{9}$                         & & $4.5550\times10^{-11}$  & & $2.4929\times10^{-9}$  & &  $9.0739\times10^{-8}$    \\
\midrule[0.5pt]
$s_1=1.9$            &  &  $D3Q19,\omega_0=\frac{1}{3},\omega_1=\frac{1}{18},\omega_7=\frac{1}{36}$&$\quad$& $2.3679\times10^{-7}$  & & $8.4546\times10^{-7}$  & &  $3.1797\times10^{-6}$    \\
                     &  &  $D3Q7,\omega_0=\frac{1}{3},\omega_1=\frac{1}{9}$                         & & $3.2065\times10^{-7}$  & & $1.2394\times10^{-6}$  & &  $4.8296\times10^{-6}$    \\
 \hline \hline
\end{tabular}
\label{Tab_Exp3_4}
\end{table}
\subsubsection{Helmholtz equation}
\begin{figure}[ht]
\centering
\subfigure[]{ \label{fig:mini:subfig:a8}
\includegraphics[scale=0.3]{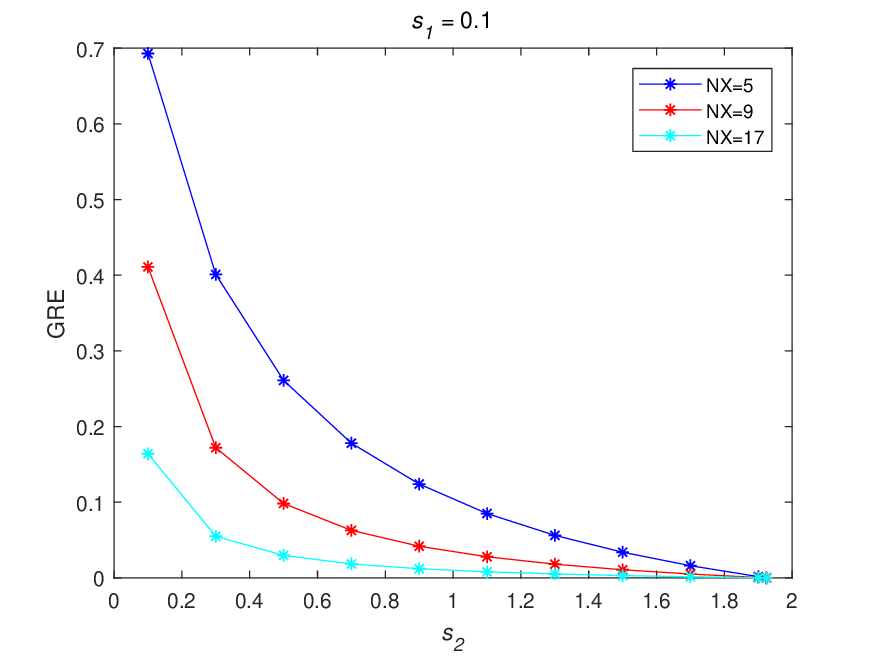}}
\subfigure[]{ \label{fig:mini:subfig:b8}
\includegraphics[scale=0.3]{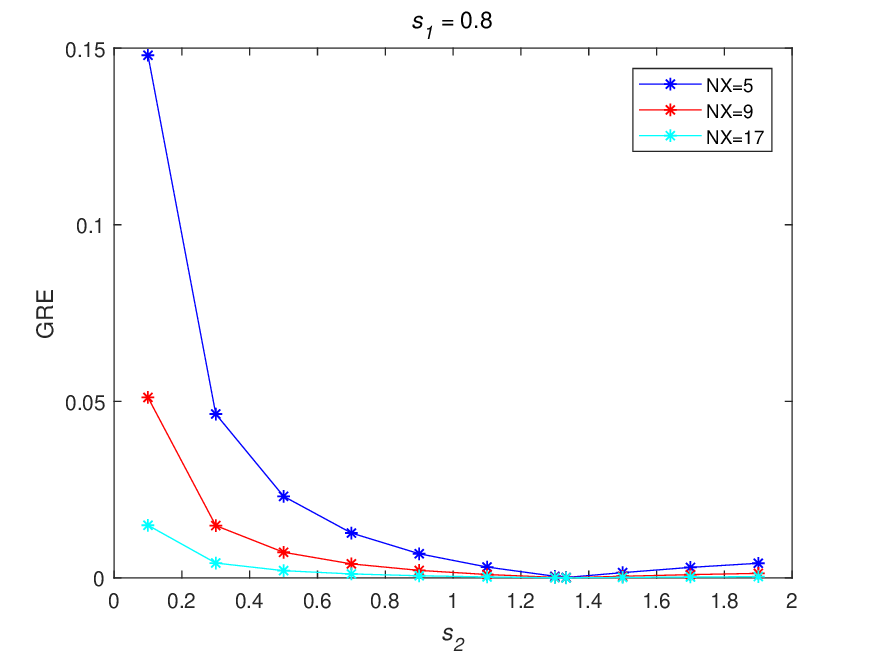}}
\subfigure[]{ \label{fig:mini:subfig:c8}
\includegraphics[scale=0.3]{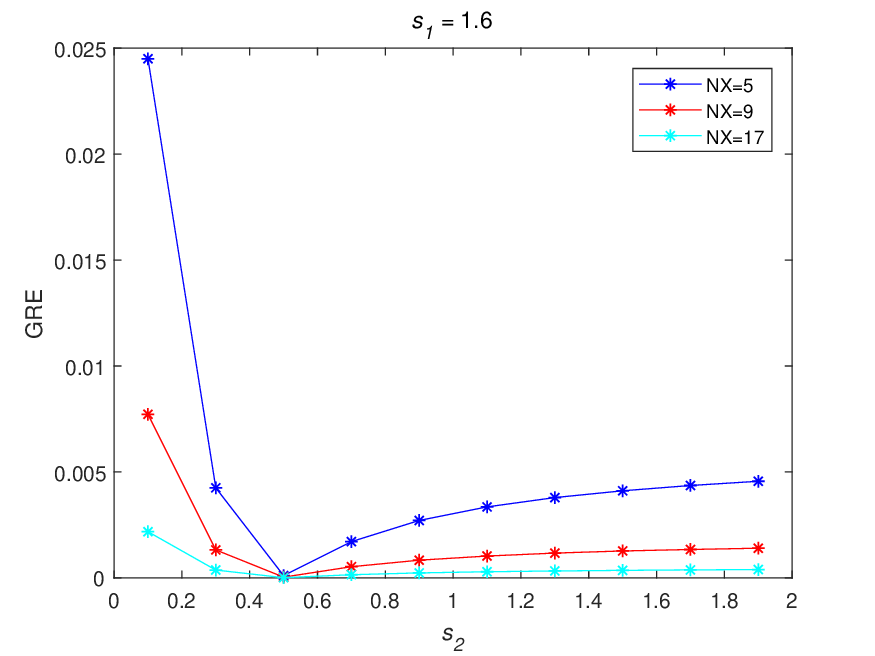}}
\caption{(Color online) The GREs of D2Q9 MRT model with ABB boundary scheme and weight coefficient $\omega_0={4}/{9}$, $\omega_1={1}/{9}$, $\omega_5={1}/{36}$.}
\label{Exp4_Fig1}
\end{figure}

\begin{figure}[ht]
\centering
\subfigure[]{ \label{fig:mini:subfig:a9}
\includegraphics[scale=0.3]{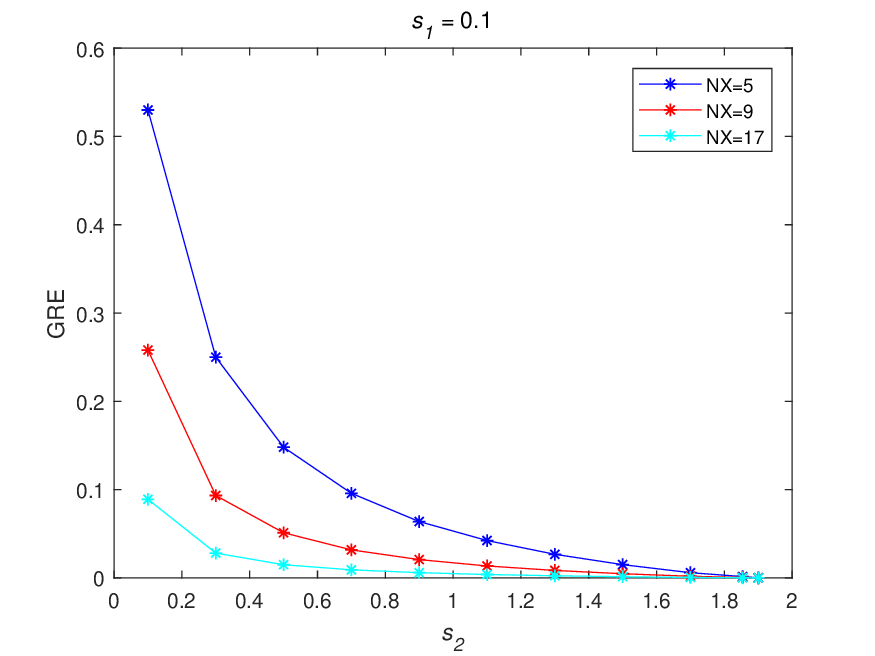}}
\subfigure[]{ \label{fig:mini:subfig:b9}
\includegraphics[scale=0.3]{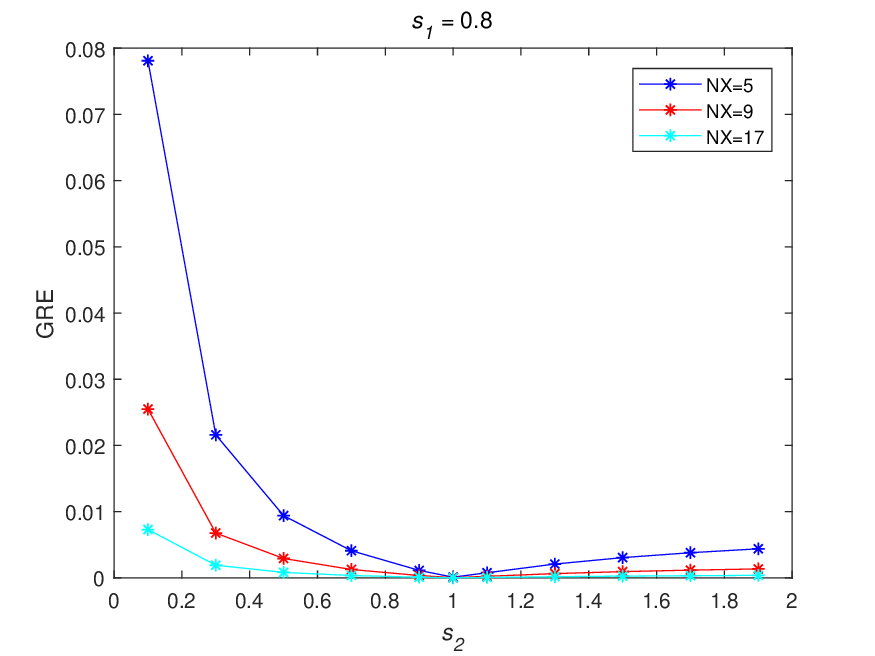}}
\subfigure[]{ \label{fig:mini:subfig:c9}
\includegraphics[scale=0.3]{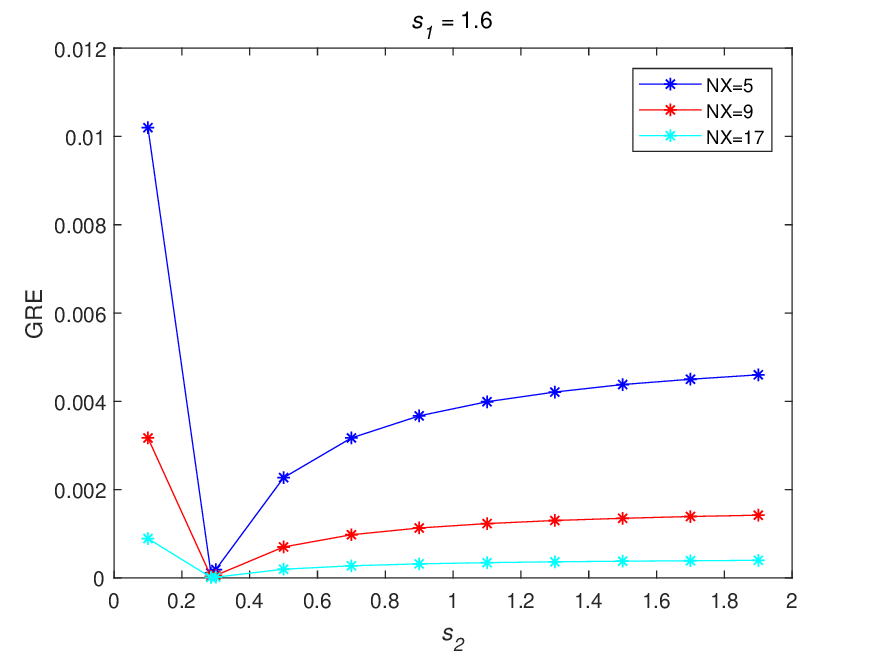}}
\caption{(Color online) The GREs of D2Q9 MRT model with ABB boundary scheme and weight coefficient $\omega_0={1}/{9}$, $\omega_1={1}/{9}$, $\omega_5={1}/{9}$.}
\label{Exp4_Fig2}
\end{figure}

\begin{figure}[ht]
\centering
\subfigure[]{ \label{fig:mini:subfig:a10}
\includegraphics[scale=0.3]{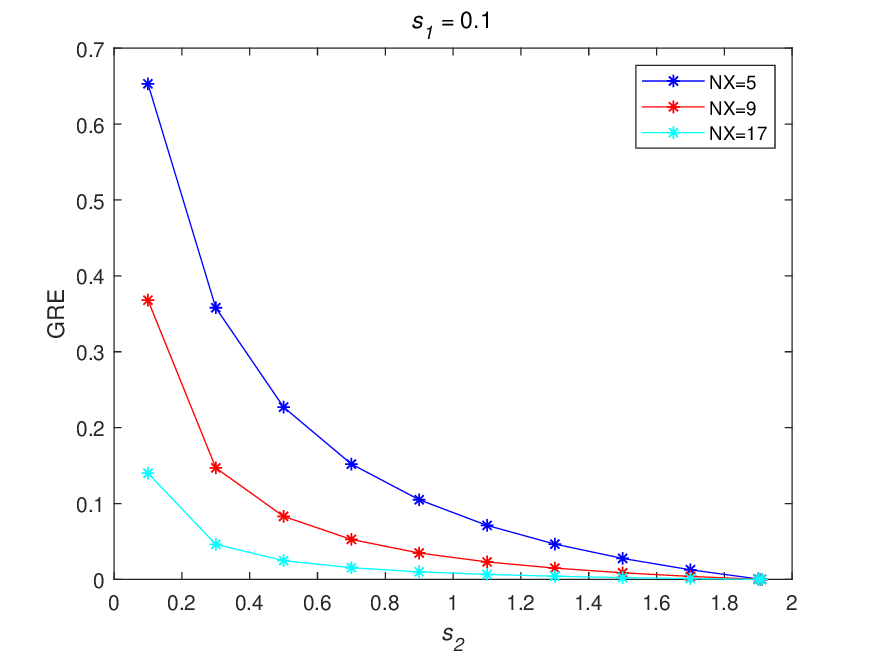}}
\subfigure[]{ \label{fig:mini:subfig:b10}
\includegraphics[scale=0.3]{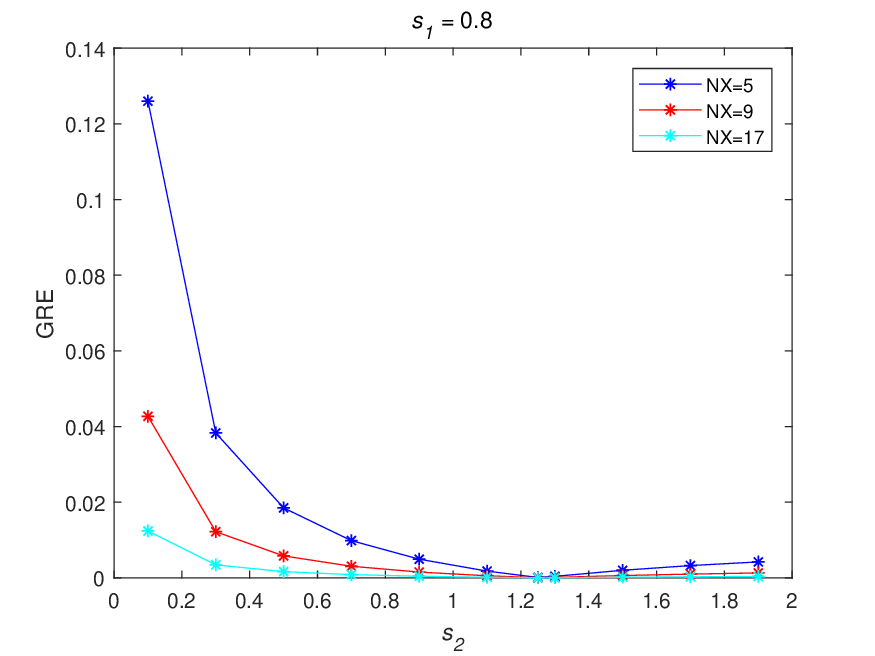}}
\subfigure[]{ \label{fig:mini:subfig:c10}
\includegraphics[scale=0.3]{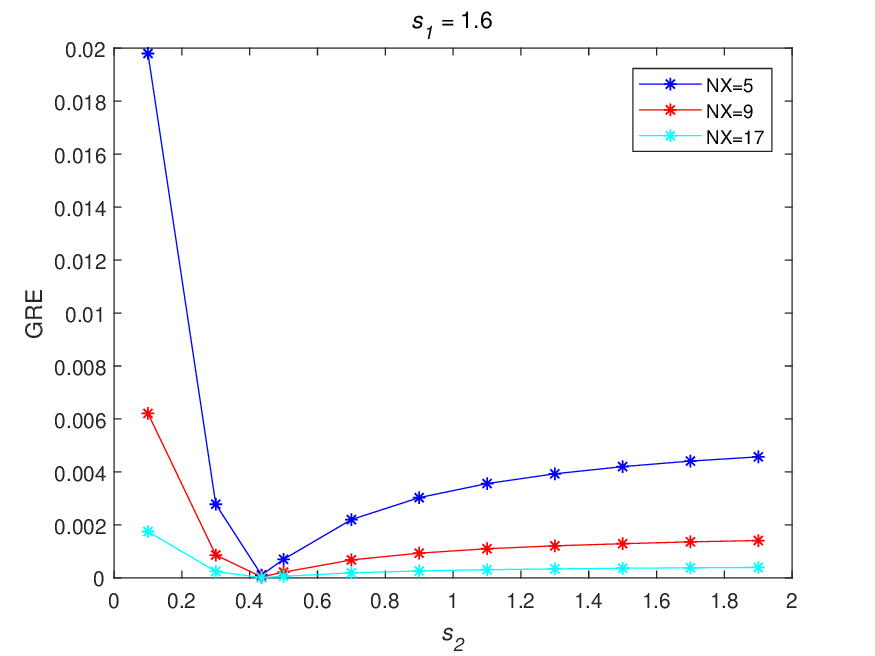}}
\caption{(Color online) The GREs of D2Q9 MRT model with ABB boundary scheme and weight coefficient $\omega_0={1}/{3}$, $\omega_1={1}/{9}$, $\omega_5={1}/{18}$.}
\label{Exp4_Fig3}
\end{figure}

We also concidered the following linear Helmholtz equation, as
\begin{equation}\label{eq65}
\frac{\partial \phi}{\partial t}=\nabla^2 \phi-(\lambda^2+\mu^2)\phi,
\end{equation}
with the boundary conditions
\begin{equation}\label{eq66}
\begin{aligned}
\phi=0,\quad &0<x<H, &y=H,\\
\phi=e^{-\lambda x},\quad &0<x<H, &y=0,\\
\phi=\frac{{\rm sinh}[\mu(1-y)]}{{\rm sinh}(\mu)},\quad &0<y<H, &x=0,\\
\lambda \phi+\frac{\partial \phi}{\partial x}=0,\quad &0<y<H, &x=H.\\
\end{aligned}
\end{equation}
The physical domain is $\Omega=[0,H]\times[0,H]$, $\lambda$ and $\mu$ are two constants.
Under above conditions, steady analytical solution of Eq. (\ref{eq65}) can be obtained
\begin{equation}\label{eq67}
\phi^{*}(x,y)=e^{-\lambda x}\frac{{\rm sinh}[\mu(1-y)]}{{\rm sinh}(\mu)},
\end{equation}
which is more complicated than Eq. (\ref{eq63}).
We conducted some simulations with $\lambda=0$ and $\mu=1.0$,
and present the results of D2Q9 MRT model under different values of $s_1$ in Figs. \ref{Exp4_Fig1}, \ref{Exp4_Fig2}, \ref{Exp4_Fig3}, where different weight coefficients are used. As we can see, the analytical solution Eq. (\ref{eq67}) is time-independent and only depends on y when $\lambda=0$ and $\mu=1.0$.
As shown in these figures, the relaxation parameter $s_2$ has a significant effect on numerical results, what is more, we can obtain the most accurate results when the value of $s_2$ determined by Eq. (\ref{eq48}) is adopted.

\subsection{A unidirectional time-independent complex-valued CDEs}
In this part, we further considered a simple two-dimension complex-valued problem governed by Eq. (\ref{eq63}) to verify Eqs. (\ref{eq48}) and (\ref{eq53}) where $D = 1+i$, $R = 4i$, $L = 1.0$, $u_x = 0.1$, $u_y = 0.0$,
and the boundary conditions $\phi_0 =0 $, $\phi_L = 1+i$.
In our simulations, $\delta_x = {L}/{N}$ with the grid number $N$ varying from 5 to 17, the D2Q5 MRT model ($\theta=0$) is used.

The $\tau_r$, $\tau_i$ are the relaxation times of the real and the imaginary parts respectively, and $S_r=diag(s_0,s_{r1},s_{r1},s_{r2},s_{r2})$ and $S_i=diag(s_0,s_{i1},s_{i1},s_{i2},s_{i2})$ are the diagonal relaxation matrix. Then we have \cite{Shi2009}
\begin{equation}\label{eq68}
\tau_r=\frac{D_r}{c_s^2\Delta t}+\frac{1}{2},\quad\tau_i=\frac{D_i}{c_s^2\Delta t},\quad s_{r1}=\frac{\tau_r}{\tau_r^2+\tau_i^2},\quad s_{i1}=-\frac{\tau_i}{\tau_r^2+\tau_i^2}.
\end{equation}
where $D=D_r+iD_i$.
In our simulations, we take $s_0$ = 0.0, $s_{r1}$ = 1.0, 10.0, 0.501, and $s_{i1}$ is determined by Eq. (\ref{eq68}).
Substituting $s_1=s_{r1}+is_{i1}$ and $s_2=s_{r2}+is_{i2}$ into Eq. (\ref{eq48}), we have
\begin{equation}\label{eq69}
s_{r2}[-4+s_{r1}+4(2-s_{r1})a_1\theta]-s_{i2}s_{i1}(1-4a_1\theta)+4(2-s_{r1})a_0=0,
\end{equation}
\begin{equation}\label{eq70}
s_{i2}[-4+s_{r1}+4(2-s_{r1})a_1\theta]+s_{r2}s_{i1}(1-4a_1\theta)-a_0s_{i1}=0,
\end{equation}
where $a_0=\omega_0+2\omega_1$, $a_1=\omega_1$ in the D2Q5 model.
The $s_{r2}$ and $s_{i2}$ are choose to satisfy Eqs. (\ref{eq69}) and (\ref{eq70}), and it shows a good accuracy in Tables. \ref{Tab_Exp6_1} and \ref{Tab_Exp6_3}.
\begin{table}[ht]
\caption{The GREs of D2Q5 MRT model for the complex cases with ABB boundary scheme
 ($\omega_0={1}/{3},\omega_1={1}/{6}$).}
\centering
\begin{tabular}{cclcccccccc}
\hline \hline
\multicolumn{4}{c}{Different models}   $\quad$  & $N=5$             & $\quad$  & $N=9$            & $\quad$ & $N=17$                  \\
\midrule[1pt]
$MRT$                &  &  $\tau_r=1.0,\tau_i=0.5$      & & $1.2775\times10^{-16}$  & & $2.0708\times10^{-9}$  & &  $1.1467\times10^{-7}$    \\
                     &  &  $\tau_r=10.0,\tau_i=9.5$        & & $4.1977\times10^{-16}$  & & $1.2100\times10^{-14}$  & &  $1.9386\times10^{-10}$    \\
                     &  &  $\tau_r=0.501,\tau_i=0.001$   & & $4.4416\times10^{-6}$  & & $1.5648\times10^{-5}$  & &  $5.8706\times10^{-5}$    \\
 \hline \hline
\end{tabular}
\label{Tab_Exp6_1}
\end{table}
\begin{table}[ht]
\caption{The $E_{max}$ of D2Q5 MRT model for the complex cases with ABB boundary scheme
 ($\omega_0={1}/{3},\omega_1={1}/{6}$).}
\centering
\begin{tabular}{cclcccccccc}
\hline \hline
\multicolumn{4}{c}{Different models}   $\quad$  & $N=5$             & $\quad$  & $N=9$            & $\quad$ & $N=17$                  \\
\midrule[1pt]
$MRT$                &  &  $\tau_r=1.0,\tau_i=0.5$      & & $1.1102\times10^{-16}$  & & $2.5946\times10^{-9}$  & &  $1.3973\times10^{-7}$    \\
                     &  &  $\tau_r=10.0,\tau_i=9.5$        & & $7.2165\times10^{-16}$  & & $1.7431\times10^{-14}$  & &  $2.3966\times10^{-10}$    \\
                     &  &  $\tau_r=0.501,\tau_i=0.001$   & & $6.4009\times10^{-6}$  & & $2.2385\times10^{-5}$  & &  $8.4811\times10^{-5}$    \\
 \hline \hline
\end{tabular}
\label{Tab_Exp6_3}
\end{table}
Then we take the same simulation with BB boundary scheme. $s_2$ is satisfied $s1+s_2=2$, that is $s_{r2}=2-s_{r1}$,
$s_{i2}=-s_{r1}$ and shows the results in Tables. \ref{Tab_Exp6_2} and \ref{Tab_Exp6_4} which have good agreement with analytical solutions.
\begin{table}[ht]
\caption{The GREs of D2Q5 MRT model for the complex cases with BB boundary scheme
 ($\omega_0={1}/{3},\omega_1={1}/{6}$).}
\centering
\begin{tabular}{cclcccccccc}
\hline \hline
\multicolumn{4}{c}{Different models}   $\quad$  & $N=5$             & $\quad$  & $N=9$            & $\quad$ & $N=17$                  \\
\midrule[1pt]
$MRT$                &  &  $\tau_r=1.0,\tau_i=0.5$      & & $3.2814\times10^{-16}$  & & $1.4988\times10^{-11}$  & &  $1.0332\times10^{-7}$    \\
                     &  &  $\tau_r=10.0,\tau_i=9.5$        & & $1.6129\times10^{-11}$  & & $5.6665\times10^{-11}$  & &  $1.0665\times10^{-9}$    \\
                     &  &  $\tau_r=0.501,\tau_i=0.001$   & & $2.8179\times10^{-6}$  & & $1.2367\times10^{-5}$  & &  $5.2043\times10^{-5}$    \\
 \hline \hline
\end{tabular}
\label{Tab_Exp6_2}
\end{table}
\begin{table}[ht]
\caption{The $E_{max}$ of D2Q5 MRT model for the complex cases with BB boundary scheme
 ($\omega_0={1}/{3},\omega_1={1}/{6}$).}
\centering
\begin{tabular}{cclcccccccc}
\hline \hline
\multicolumn{4}{c}{Different models}   $\quad$  & $N=5$             & $\quad$  & $N=9$            & $\quad$ & $N=17$                  \\
\midrule[1pt]
$MRT$                &  &  $\tau_r=1.0,\tau_i=0.5$      & & $4.4409\times10^{-16}$  & & $2.3118\times10^{-11}$  & &  $1.4393\times10^{-7}$    \\
                     &  &  $\tau_r=10.0,\tau_i=9.5$        & & $2.2474\times10^{-11}$  & & $7.3081\times10^{-11}$  & &  $1.6065\times10^{-9}$    \\
                     &  &  $\tau_r=0.501,\tau_i=0.001$   & & $4.4960\times10^{-6}$  & & $1.8838\times10^{-5}$  & &  $7.7969\times10^{-5}$    \\
 \hline \hline
\end{tabular}
\label{Tab_Exp6_4}
\end{table}
\section{CONCLUSIONS}
In this work, we performed a detailed analysis on the discrete effects of ABB, BB and NEE schemes of the popular one- to three- dimensional D$n$Q$q$  MRT LB model for real- and complex-valued CDEs.
Firstly, through the analysis with ABB boundary scheme, we obtain a relation with four adjustable parameters the weight coefficient, the relaxation factors $s_1$ and $s_2$ associated with first and second moments and a model parameter $\theta$, which can be used to eliminate the discrete effect.
We would also like to point out that taking $\theta=1$ under some assumption, the relation in \cite{Ginzburg2017} in the framework of TRT model would be the special case of Eq. (\ref{eq48}).
The weight coefficient $\omega$ can be considered as an adjustable parameter makes the general relation Eq. (\ref{eq48}) more flexible. Then we analyse the discrete effects of BB and NEE boundary schemes and indicate that the discrete effect of BB scheme can be eliminated when $s_1+s_2=2$, and the discrete effect of NEE scheme can not be eliminated except $s_1=1$. The adjustment of ABB boundary scheme is more flexible than BB and NEE boundary schemes.
We also carried out some numerical simulations of several special equations, including the real-valued linear time-independent diffusion equations in two- and three-dimensional space, the real-valued two-dimensional Helmholtz equation, and the complex-valued linear time-independent diffusion equation. The results also show that when the relation Eq. (\ref{eq48}) for ABB boundary scheme and $s_1+s_2=2$ for BB boundary scheme is satisfied, the discrete effect (or numerical slip) can be eliminated.
\section*{Acknowledgements}
This work is supported by the National Natural Science Foundation of China (Grants No. 51576079 and No.
51836003), and the National Key Research and Development Program of China (Grant No. 2017YFE0100100)
\section*{APPENDIX}
\subsection{Equivalent difference equation of the MRT model}
In this Appendix, we show how to derive the equivalent difference equation.
Firstly, for the D1Q3 MRT model, from Eq. (\ref{eq13}), we can obtain the expressions of the distribution functions,
\begin{subequations}
    \begin{equation}\label{eq71:a}
    f^{k-1}_{-1}=f^{k}_{-1}-(\frac{s_1}{2}+\frac{s_2}{2})(f^k_{-1}-f^{k,eq}_{-1})-(\frac{s_2}{2}-\frac{s_1}{2})(f^k_{1}-f^{k,eq}_{1})+w_1(1-\frac{\theta s_2}{2})\delta_tR,
    \end{equation}
    \begin{equation}\label{eq71:b}
    f^{k}_{0}=f^{k}_{0}-(s_0-s_2)(f^k_{-1}-f^{k,eq}_{-1})-s_0(f^k_{0}-f^{k,eq}_{0})-(s_0-s_2)(f^k_{1}-f^{k,eq}_{1})+[w_1\theta (s_2-s_0)
    +w_0(1-\frac{\theta s_0}{2})]\delta_tR,
    \end{equation}
    \begin{equation}\label{eq71:c}
    f^{k+1}_{1}=f^{k}_{1}-(\frac{s_2}{2}-\frac{s_1}{2})(f^k_{-1}-f^{k,eq}_{-1})-(\frac{s_2}{2}+\frac{s_1}{2})(f^k_{1}-f^{k,eq}_{1})+w_1(1-\frac{\theta s_2}{2})\delta_tR,
    \end{equation}
\end{subequations}
where $f^k_i$, $f^{k,eq}_i$ are the distribution function and its equilibrium part at $x = k\delta_x$.
According to Eqs. (\ref{eq12}) and (\ref{eq3}), we have
\begin{equation}\label{eq72}
\phi_k =f_{-1}^k+f_0^k+f_1^k+\frac{\theta R}{2}\delta t,
\end{equation}
\begin{equation}\label{eq73}
f_0^{k,eq}=\omega_0\phi_k,\quad  f_1^{k,eq}=\omega_1\phi_k+\frac{u_k\phi_k}{2c},\quad  f_{-1}^{k,eq}=\omega_1\phi_k-\frac{u_k\phi_k}{2c}.
\end{equation}
Substituting Eq. (\ref{eq72}) into Eq. (\ref{eq71:b}), one can obtain
\begin{equation}\label{eq74}
f^k_{-1}+f^k_1=\phi_k-f^{k,eq}_0+A\delta_tR,\quad A=-\frac{\theta }{2}-\frac{s_2-s_0}{s_2}\theta(\omega_1-\frac{1}{2})-\omega_0(\frac{1}{s_2}-\frac{\theta s_0}{2s_2}).
\end{equation}
Based on Eq. (\ref{eq74}), we can get
\begin{subequations}
    \begin{equation}\label{eq75:a}
    f^k_{-1}=\phi_k-f^{k,eq}_0+A\delta_tR-f^k_1,
    \end{equation}
    \begin{equation}\label{eq75:b}
    f^k_{1}=\phi_k-f^{k,eq}_0+A\delta_tR-f^k_{-1}.
    \end{equation}
\end{subequations}
Substituting Eq. (\ref{eq75:a}) into Eq. (\ref{eq71:c}), and with the help of Eq. (\ref{eq74}), we have
\begin{equation}\label{eq76}
f^{k+1}_{1}=(1-s_1)f^k_1+s_1f_1^{k,eq}+B\delta_tR, \quad B=\omega_1(1-\frac{\theta s_2}{2})-(\frac{s_2}{2}-\frac{s_1}{2})A.
\end{equation}
Similarly, substituting Eq. (\ref{eq75:b}) into Eq. (\ref{eq71:a}), and with the aid of Eq. (\ref{eq74}), one can obtain
\begin{equation}\label{eq77}
f^{k-1}_{-1}=(1-s_1)f^k_{-1}+s_1f_{-1}^{k,eq}+B\delta_tR.
\end{equation}
In addition, from Eqs. (\ref{eq76}) and (\ref{eq77}), we also have
\begin{subequations}
    \begin{equation}\label{eq78:a}
    f^{k}_{1}=(1-s_1)f^{k-1}_1+s_1f_1^{k-1,eq}+B\delta_tR,
    \end{equation}
    \begin{equation}\label{eq78:b}
    f^{k}_{-1}=(1-s_1)f^{k+1}_{-1}+s_1f_{-1}^{k+1,eq}+B\delta_tR.
    \end{equation}
\end{subequations}
Summing Eqs. (\ref{eq78:a}) and (\ref{eq78:b}), one can derive the following equation,
\begin{equation}\label{eq79}
\begin{aligned}
f^{k}_{1}+f^{k}_{-1}=&(1-s_1)[2\omega_1(\phi_{k+1}+\phi_{k-1})-s_1(f^{k,eq}_{-1}+f^{k,eq}_{1})-(1-s_1)(f_1^k+f_{-1}^k)+2(A-B)\delta_tR]\\
&+s_1(f_{-1}^{k+1,eq}+f_{1}^{k-1,eq})+2B\delta_tR, \\
\end{aligned}
\end{equation}
where Eqs. (\ref{eq76}) and (\ref{eq77}) have been used.
Substituting Eq. (\ref{eq74}) into Eq. (\ref{eq79}) yields
\begin{equation}\label{eq80}
\omega_1\frac{s_1-2}{s_1}(\phi_{k+1}+\phi_{k-1}-2\phi_k)=\frac{\phi_{k+1}u_{k+1}-\phi_{k-1}u_{k-1}}{2c}+\delta_tR,
\end{equation}
where Eq. (\ref{eq73}) has been adopted. From Eq. (\ref{eq80}), we can obtain the equivalent difference equation of the MRT model, i.e., Eq. (\ref{eq15}).

For the D3Q27 model, we have
\begin{subequations}
    \begin{equation}\label{eq81:a}
    \begin{aligned}
    f^{k-1}_{4,7,10,11,12,19,20,23,24}=&f^{k}_{4,7,10,11,12,19,20,23,24}-(\frac{s_1}{2}+\frac{s_2}{2})(f^k_{4,7,10,11,12,19,20,23,24}-f^{k,eq}_{4,7,10,11,12,19,20,23,24})\\
    &-(\frac{s_2}{2}-\frac{s_1}{2})(f^k_{2,8,9,13,14,21,22,25,26}-f^{k,eq}_{2,8,9,13,14,21,22,25,26})+(\omega_1+2\omega_5)(1-\frac{\theta s_2}{2})\delta_tR,
    \end{aligned}
    \end{equation}
    \begin{equation}\label{eq81:b}
    \begin{aligned}
    f^{k}_{0,1,3,5,6,15,16,17,18}=&f^{k}_{0,1,3,5,6,15,16,17,18}-(s_0-s_2)(f^k_{4,7,10,11,12,19,20,23,24}-f^{k,eq}_{4,7,10,11,12,19,20,23,24})\\
    &-s_0(f^k_{013}-f^{k,eq}_{013})-(s_0-s_2)(f^k_{2,8,9,13,14,21,22,25,26}-f^{k,eq}_{2,8,9,13,14,21,22,25,26})\\
    &+[(\omega_1+2\omega_5)\theta (s_2-s_0)+(\omega_0+2\omega_1)(1-\frac{\theta s_0}{2})]\delta_tR, \\
    \end{aligned}
    \end{equation}
    \begin{equation}\label{eq81:c}
    \begin{aligned}
    f^{k+1}_{2,8,9,13,14,21,22,25,26}=&f^{k}_{2,8,9,13,14,21,22,25,26}-(\frac{s_2}{2}-\frac{s_1}{2})(f^k_{4,7,10,11,12,19,20,23,24}-f^{k,eq}_{4,7,10,11,12,19,20,23,24})\\
    &-(\frac{s_2}{2}+\frac{s_1}{2})(f^k_{2,8,9,13,14,21,22,25,26}-f^{k,eq}_{2,8,9,13,14,21,22,25,26})+(\omega_1+2\omega_5)(1-\frac{\theta s_2}{2})\delta_tR,
    \end{aligned}
    \end{equation}
\end{subequations}
where $f^k_{i,j,m}=f^k_i+f^k_j+f^k_m$, $f^{k,eq}_{i,j,m}=f^{k,eq}_i+f^{k,eq}_j+f^{k,eq}_m$.
If the parts of $f^k_{0,1,3,5,6,15,16,17,18}$, $f^k_{2,8,9,13,14,21,22,25,26}$, and $f^k_{4,7,10,11,12,19,20,23,24}$ in the D3Q27 model are viewed as $f^k_0$, $f^k_1$, and $f^k_{-1}$ in the D1Q3 model, $w_0+4w_1+4w_7$ and $w_1+4w_7+4w_{19}$ in the D3Q27 model are considered as $w_0$ and $w_1$ in D1Q3 model, we can derive the equivalent different Eq. (\ref{eq15}) through the similar process.
\subsection{Discrete effect of the ABB boundary condition}
In the D1Q3 model, when $k=1$, $k=N$, Eqs. (\ref{eq78:b}) and (\ref{eq78:a}) can be written as
\begin{subequations}
    \begin{equation}\label{eq82:a}
    f^{1}_{-1}=(1-s_1)f^2_{-1}+s_1f_{-1}^{2,eq}+B\delta_tR.
    \end{equation}
    \begin{equation}\label{eq82:b}
    f^{N}_{1}=(1-s_1)f^{N-1}_{1}+s_1f_{1}^{N-1,eq}+B\delta_tR.
    \end{equation}
\end{subequations}
Substituting Eq. (\ref{eq75:a}) into Eq. (\ref{eq82:a}),  substituting Eq. (\ref{eq75:b}) into Eq. (\ref{eq82:b}), we can obtain
\begin{subequations}
    \begin{equation}\label{eq83:a}
    f^{1}_{-1}=(1-s_1)(\phi_2-f^{2,eq}_0+A\delta_tR-f^2_1)+s_1f_{-1}^{2,eq}+B\delta_tR.
    \end{equation}
    \begin{equation}\label{eq83:b}
    f^{N}_{1}=(1-s_1)(\phi_{N-1}-f^{N-1,eq}_0+A\delta_tR-f^{N-1}_{-1})+s_1f_{1}^{N-1,eq}+B\delta_tR.
    \end{equation}
\end{subequations}
In addition, substituting Eqs. (\ref{eq73}) and (\ref{eq78:a}) into Eqs. (\ref{eq83:a}) and (\ref{eq83:b}) gives rise to
\begin{subequations}
    \begin{equation}\label{eq84:a}
    s_1f^{1}_{-1}=\omega_1\phi_2+(s_1-1)\omega_1\phi_1+\frac{(s_1A-B)(1-s_1)+B}{2-s_1}\delta_tR.
    \end{equation}
    \begin{equation}\label{eq84:b}
    s_1f^{N}_{1}=\omega_1\phi_{N-1}+(s_1-1)\omega_1\phi_N+\frac{(s_1A-B)(1-s_1)+B}{2-s_1}\delta_tR.
    \end{equation}
\end{subequations}
On the other hand, the ABB scheme can be given by
\begin{subequations}
    \begin{equation}\label{eq85:a}
    f^1_{1}=-f^{1,+}_{-1}+2\omega_1\phi_0.
    \end{equation}
    \begin{equation}\label{eq85:b}
    f^N_{-1}=-f^{N,+}_{1}+2\omega_1\phi_L.
    \end{equation}
\end{subequations}
Substituting Eq. (\ref{eq78:b}) into Eq. (\ref{eq85:a}), and  substitute Eq. (\ref{eq78:a}) into Eq. (\ref{eq85:b}), one can obtain
\begin{subequations}
    \begin{equation}\label{eq86:a}
    f^1_{1}=-[(1-s_1)f^1_{-1}+s_1f_{-1}^{1,eq}+B\delta_tR]+2\omega_1\phi_0,
    \end{equation}
    \begin{equation}\label{eq86:b}
    f^N_{-1}=-[(1-s_1)f^N_{1}+s_1f_{1}^{N,eq}+B\delta_tR]+2\omega_1\phi_L.
    \end{equation}
\end{subequations}
Substituting Eqs. (\ref{eq85:a}) and (\ref{eq85:b}) into Eqs. (\ref{eq86:a}) and (\ref{eq86:b}), we can obtain
\begin{subequations}
    \begin{equation}\label{eq87:a}
    \omega_1(-\phi_2+3\phi_1-2\phi_0)=[\frac{(s_1A-B)(1-s_1)+B}{2-s_1}-A-B]\delta_tR,
    \end{equation}
    \begin{equation}\label{eq87:b}
    \omega_1(-\phi_{N-1}+3\phi_N-2\phi_L)=[\frac{(s_1A-B)(1-s_1)+B}{2-s_1}-A-B]\delta_tR,
    \end{equation}
\end{subequations}
which can also be written as
\begin{subequations}
    \begin{equation}\label{eq88:a}
    \omega_1(-\phi_2+3\phi_1-2\phi_0)=\frac{-2+s_1+s_2-s_1s_2+w_1(s_1-2)(s_2-2)}{s_2(s_1-2)}\delta_tR,
    \end{equation}
    \begin{equation}\label{eq88:b}
    \omega_1(-\phi_{N-1}+3\phi_N-2\phi_L)=\frac{-2+s_1+s_2-s_1s_2+w_1(s_1-2)(s_2-2)}{s_2(s_1-2)}\delta_tR.
    \end{equation}
\end{subequations}
From Eq. (\ref{eq36}), we have
\begin{subequations}
    \begin{equation}\label{eq89:a}
    \phi_1=-\frac{\Delta \phi}{N^2}+(2N+1)\frac{\Delta \phi}{N^2}-(4N+1)\frac{\Delta \phi}{N^2}+\frac{1}{2}(\phi_s^{N+0.5}-\phi_s^{0.5})+\phi_0+\phi_s^{0.5},
    \end{equation}
    \begin{equation}\label{eq89:b}
    \phi_2=-\frac{4\Delta \phi}{N^2}+(2N+1)\frac{2\Delta \phi}{N^2}-(4N+1)\frac{\Delta \phi}{N^2}+\frac{3}{2}(\phi_s^{N+0.5}-\phi_s^{0.5})+\phi_0+\phi_s^{0.5},
    \end{equation}
    \begin{equation}\label{eq89:c}
    \phi_{N-1}=-\frac{\Delta \phi}{N^2}(N-1)^2+(2N+1)\frac{\Delta \phi}{N^2}(N-1)-(4N+1)\frac{\Delta \phi}{N^2}+(N-\frac{3}{2})(\phi_s^{N+0.5}-\phi_s^{0.5})+\phi_0+\phi_s^{0.5},
    \end{equation}
    \begin{equation}\label{eq89:d}
    \phi_N=-\Delta \phi+(2N+1)\frac{\Delta \phi}{N}-(4N+1)\frac{\Delta \phi}{N^2}+(N-\frac{1}{2})(\phi_s^{N+0.5}-\phi_s^{0.5})+\phi_0+\phi_s^{0.5}.
    \end{equation}
\end{subequations}
Substituting Eqs. (\ref{eq89:a}) and (\ref{eq89:b}) into Eq. (\ref{eq88:a}), and Eqs. (\ref{eq89:c}) and (\ref{eq89:d}) into Eq. (\ref{eq88:b}), we can obtain Eqs. (\ref{eq39}) and (\ref{eq40}).

\section{References}
\bibliographystyle{elsarticle-num}
\bibliography{ref}
\end{document}